\documentclass[preprint, 5p, times, twocolumn]{IEEEtran}
\usepackage[colorlinks, 
            linkcolor=red, 
            anchorcolor=blue, 
            citecolor=green]{hyperref}
\usepackage{amssymb}
\usepackage{amsmath}
\usepackage{cleveref}
\crefname{figure}{fig.}{figures}
\Crefname{figure}{Fig.}{Figures}
\usepackage{CJKutf8}
\usepackage{graphicx} 
\usepackage{float} 
\usepackage{subfig} 
\usepackage{booktabs}
\usepackage{stfloats}

\usepackage{algorithm}
\usepackage{algpseudocode}
\usepackage{booktabs}
\usepackage[table,xcdraw]{xcolor}
\usepackage{graphicx}

\renewcommand{\figurename}{Fig}
\renewcommand{\thesubsubsection}{\Alph{subsubsection}}

\begin{document}

\title{A Novel Discrete Memristor-Coupled Heterogeneous Dual-Neuron Model and Its Application in Multi-Scenario Image Encryption}

\author{Yi Zou,
        Mengjiao Wang,
        Xinan Zhang,~\IEEEmembership{Senior Member, IEEE,}
        Herbert Ho-Ching Iu,~\IEEEmembership{Fellow,~IEEE}

\thanks{This work was supported in part by the Research Foundation of Education Department of Hunan Province, China(Grant No. 24A0124), the Natural Science Foundation of Hunan Province, China(Grant No. 2025JJ50391) and  the National Natural Science Foundation of China (Grant No. 62071411). (Corresponding author: Mengjiao Wang)
}
\thanks{Yi Zou and Mengjiao Wang is with the School of Computer Science, Xiangtan University, Xiangtan, Hunan 411105, China (e-mail: 202205570112@smail.xtu.edu.cn; wangmj@xtu.edu.cn) \\Xinan Zhang and Herbert Ho-Ching Iu are with the School of Electrical, Electronic and Computer Engineering, University of Western Australia, Crawley, WA 6009, Australia (e-mail: xinan.zhang@uwa.edu.au; herbert.iu@uwa.edu.au).}}

\markboth{IEEE INTERNET OF THINGS JOURNAL}%
{Shell \MakeLowercase{\textit{et al.}}: A Sample Article Using IEEEtran.cls for IEEE Journals}

\IEEEpubid{}

\maketitle

\begin{abstract}
Simulating brain functions using neural networks is a crucial area of research. Recently, the use of discrete memristor-coupled neurons has gained significant attention, as memristors function effectively as synapses, which are key components for learning and memory, emphasizing the neurobiological relevance of this model. This study proposes a discrete memristive heterogeneous dual-neuron network (MHDNN). The stability of MHDNN is analyzed, depending on initial conditions and diverse neuronal parameters. Extensive numerical analysis reveals its complex dynamical behaviors. Additionally, this study investigates various neuronal firing patterns that emerge under different coupling strengths and explores the synchronization phenomena occurring between neurons. The MHDNN  is implemented and validated on the STM32 hardware platform. An MHDNN-based image encryption algorithm is proposed, accompanied by two hardware platforms tailored for multi-scenario police image encryption. These methods ensure real-time, secure transmission of police data in complex environments, preventing hacking and enhancing system security. 

\end{abstract}


\section{Introduction}
\IEEEPARstart{T}{he} 
\label{Introduction}
 human brain is a neural network composed of neurons interconnected by numerous synapses \cite{ison2015rapid, zyarah2020neuromorphic,ma2017review}. Neurons act as fundamental units of information processing, receiving input signals, performing nonlinear transformations, and transmitting results to subsequent neurons \cite{hodgkin1952quantitative}. Understanding brain functionality has long been a central focus of scientific inquiry. In recent decades, researchers have simulated the complex activities of biological neural networks, explored the nonlinear dynamic characteristics and information processing mechanisms of the brain, and studied biologically inspired firing mechanisms \cite{chen2020bifurcation, chesebro2023ion}. To date, various neurons and neural networks have been constructed, encompassing both continuous and discrete types. Zhang et al. \cite{zhang2024bionic} were the first to utilize a memristor as a synapse to connect two heterogeneous Hopfield neural networks representing different brain regions, highlighting the advantages of high-dimensional, strongly coupled structures and complex dynamics. Xu et al. \cite{xu2023extreme} constructed a five-dimensional dual-neuron network by coupling heterogeneous Rulkov neurons with memristors, exploring the electromagnetic induction effects of memristors and investigating phase synchronization influenced by coupling strength and initial values. Wang et al. \cite{wang2020delay} examined the synchronization of two coupled memristive hyperbolic Hopfield neural networks under conditions of no delay, single delay, and multiple delays. Lai et al. \cite{lai2022generating} proposed a novel flux-controlled memristor model based on hyperbolic functions and employed it as a synapse in Hopfield neural networks, constructing three memristive HNNs capable of generating multi-double-scroll chaotic attractors or grid multi-double-scroll chaotic attractors, with the number of double-scrolls modulated by the memristor.

\vspace{-1mm}
However, the neural networks discussed above are all continuous. Discrete neural networks provide several distinct advantages over their continuous counterparts, including higher computational efficiency, stronger robustness, lower memory and energy requirements, simplified hardware implementation, and superior performance in handling discrete tasks. These characteristics make discrete neural networks especially beneficial in application scenarios such as digital hardware, embedded systems, logical reasoning, and classification \cite{indiveri2013integration, schuman2017survey, majhi2022dynamics}. Consequently, the construction of discrete neurons and neural networks has become a prominent research focus. 
Li et al. \cite{li2023offset} proposed a memristive Rulkov neuron featuring local amplitude control and coexisting homogeneous and heterogeneous multistability, demonstrating exceptional performance in secure optical fiber communication. Zhou et al. \cite{zhou2024coexisting} developed a discrete memristive-coupled Chialvo neuron that exhibited complex dynamic behaviors, including hyperchaos and multistability. Wouapi et al. \cite{wouapi2021complex} investigated the dynamics and optimal synchronization of the Hindmarsh-Rose neuron model influenced by the magnetic flux effect, addressing synchronization issues between two such neuron models. Shang et al. \cite{shang2023spatial} examined chimera states in a discrete memristive ring neural network, achieving neuron-to-neuron information exchange and showcasing the spatial morphology characteristics of biological neurons within the network.\par
Chaotic systems, characterized by deterministic nonlinearity, sensitivity to initial conditions, and long-term unpredictability, are widely used in information encryption, such as image encryption \cite{lai2023novel, li2021fractional, feng2023exploiting, wen2024cryptanalysis, ye2024visual, wang2025memristive}, secure communication \cite{liu2022secure, kekha2022robust}, and random number generation \cite{liu2023parallel, bao2023memristive}. Unlike pure randomness, chaos arises from deterministic rules, making it useful for cryptographic applications while maintaining reproducibility under identical conditions. In previous research, efforts have been made to design various chaotic encryption schemes to protect data from unauthorized access, tampering, and leakage, ensuring the confidentiality, integrity, and authenticity of information. However, with the large-scale deployment of IoT devices, they face increasing security threats such as eavesdropping, data tampering, and identity spoofing. As a result, effective encryption techniques are required to ensure the security and privacy of IoT ecosystems. The application of chaotic neural networks in IoT systems is therefore of great significance \cite{lin2024diversified,lin2022brain}.\par
Based on the above analysis, MHDNN is proposed, which exhibits rich dynamic behaviors such as hyperchaotic, chaotic, periodic, and quasi-periodic states, along with various bursting and spiking firing patterns. The main contributions of this work are as follows:
  \begin{itemize}
 \item A discrete memristive heterogeneous dual-neuron network (MHDNN) is constructed by employing a discrete tanh memristor as a synaptic coupling element between two heterogeneous one-dimensional neurons. Stability analysis is conducted to explore fixed points, revealing an infinite number of unstable points and bifurcations at critical points, which facilitate the emergence of complex dynamic behaviors.
\item We employed various numerical analysis methods, such as attractors, Lyapunov exponents, and basins of attraction, to investigate the complex dynamics of MHDNN. Additionally, we uncovered its diverse firing behaviors and the synchronization phenomenon between the two neurons.
\item A novel encryption method is proposed, combining the S-box with hyperchaotic sequences generated by MHDNN. The encryption algorithm demonstrates strong security and robustness. Additionally, two hardware platform solutions for multi-scenario image/video encryption are introduced for the first time, addressing the diverse needs of law enforcement personnel in different scenarios.
  \end{itemize}

\vspace{-2mm}
The remainder of this paper is organized as follows: The design of the tanh memristor model and MHDNN is presented in Section \eqref{sec2}. Section \eqref{sec3} investigates complex dynamical behaviors, revealing diverse firing patterns and synchronization phenomena.  Section\eqref{sec4} designs a PRNG that passes the NIST tests and develops an STM hardware platform. Additionally, two hardware platform solutions for multi-scenario image/video encryption are proposed. Finally, Section \eqref{sec5} concludes the paper.

\section{Heterogeneous two-neuron discrete model}
\label{sec2}
\subsection{Discrete memristor model}
\begin{figure}[h]
\centering
\includegraphics[width=\linewidth]{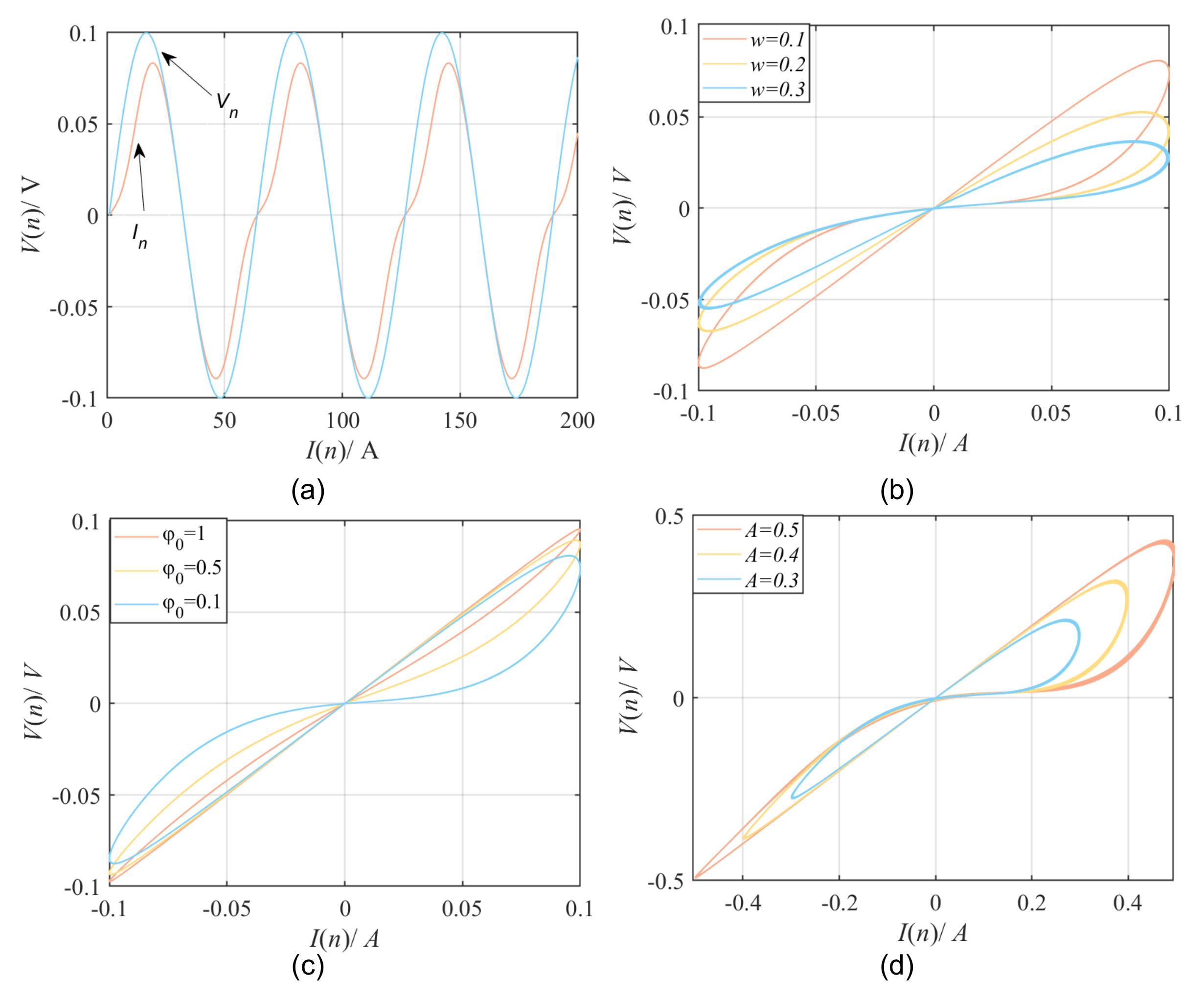}
\caption{Hysteresis loops of a discrete memristor with sinusoidal input $I_n = A\sin(\omega n)$. (a) Voltage and current sequences; (b) $A$=0.1, $\varphi_0$=0.1; (c) $A$=0.1, $\omega$=0.1; (d) $\omega$=0.3, $\varphi_0$=0.1.}
\label{Fig.1}
\end{figure}

Inspired by \cite{li2023offset}, and following the forward Euler difference theory, we construct a discrete flux-controlled memristor model with the following formulation:
\begin{equation}\begin{aligned}&V_n=W(\varphi_n)I_n=\tanh(\varphi_n)I_n\\&\varphi_{n+1}=\varphi_n+I_n\end{aligned}\label{1}\end{equation}
where \( V_n \), \( I_n \), and \( \varphi_n \) represent the voltage, current, and flux at the \( n \)-th iteration, respectively, and \( \varphi_{n+1} \) is the flux at the \( (n+1) \)-th iteration. To verify the three fingerprint features of the proposed discrete memristor, we apply \( I_n = A\sin(\omega n) \) as input. Fixing \( a = 0.1 \), \( \omega = 0.1 \), and \( I = 0.2 \), the iterative sequences of \( V_n \) and \( I_n \) are shown in \Cref{Fig.1}(a). With \( A = 0.1 \) and \( \varphi_0 = 0.1 \) fixed, the \( V_n - I_n \) curve varies with frequency, as illustrated in \Cref{Fig.1}(b). Fixing \( A = 0.1 \) and \( \omega = 0.1 \), the \( V_n - I_n \) curves under different initial magnetic values are shown in \Cref{Fig.1}(c). Similarly, with \( \omega = 0.3 \) and \( \varphi_0 = 0.1 \) fixed, the \( V_n - I_n \) curves as a function of different amplitudes are presented in \Cref{Fig.1}(d). Therefore, the numerical simulation results in \Cref{Fig.1} confirm the three fingerprint characteristics of the discrete memristor.

\subsection{Chialvo  and Rulkov neuron}
\begin{figure}[h]
\centering
\includegraphics[width=\linewidth]{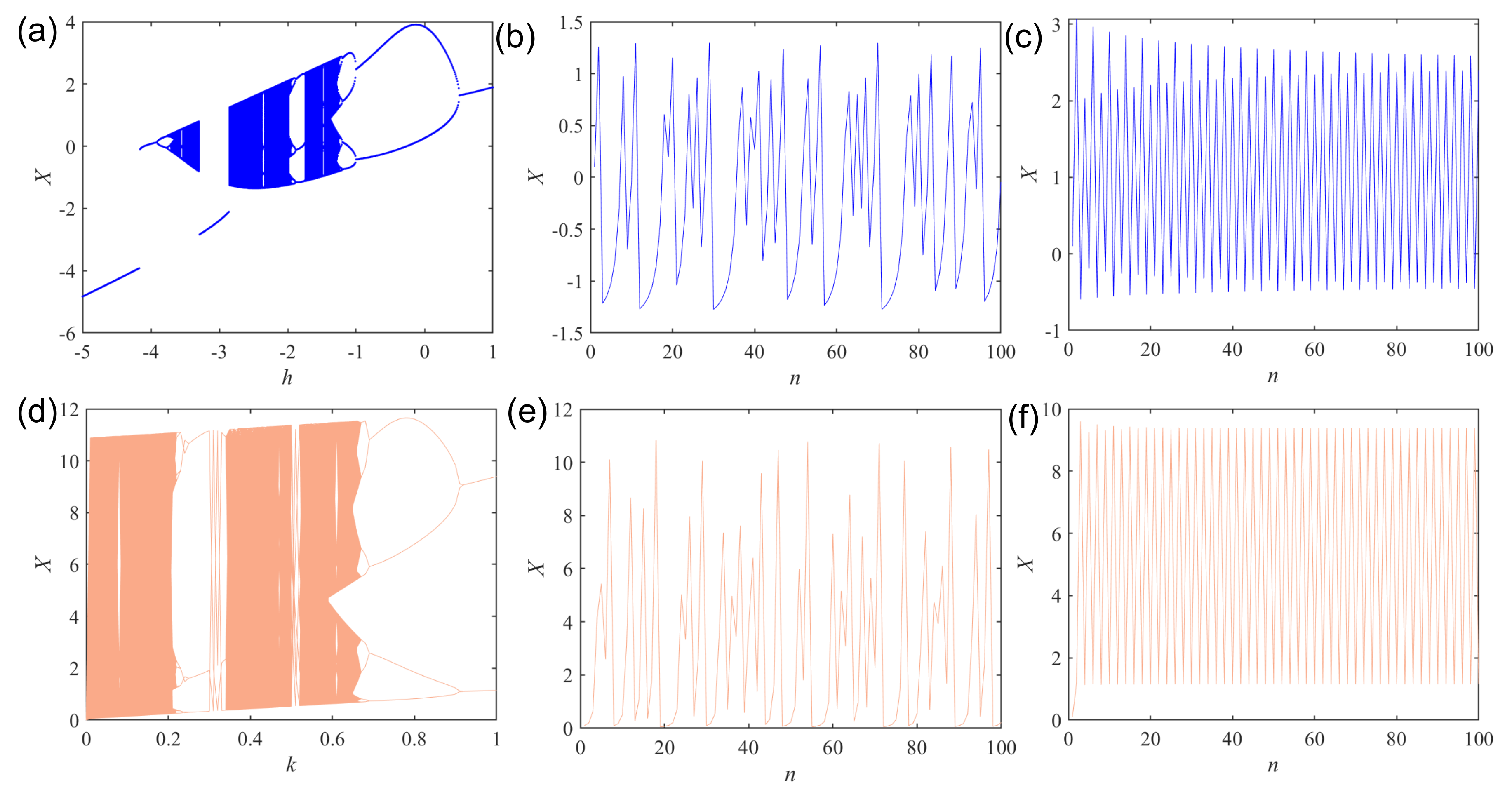} 
\caption{(a) Bifurcation diagram of the Rulkov model with fixed parameter $a$=4.1 and $ h \in [-5, 1] $; (b) Chaotic spiking when $a$=4.1 and $h$=-2.8; (c) Chaotic spiking when $a$=4.1 and $h$ = 0.99; (d) Bifurcation diagram of the Chialvo model with fixed parameters $b$=3, $c$=1, and $ k \in [0, 1] $; (e) Chaotic spiking when $b$=3, $c$ = 1, and $k$ = 0.01; (f) Periodic spiking when $b$=3, $c$=1, and $k$ = 1.}
\label{Fig.2}
\end{figure}

As one of the classic discrete neuron models, the Chialvo \cite{chialvo1995generic} and Rulkov \cite{rulkov2001regularization} neurons can exhibit diverse dynamic behaviors, such as spiking, bursting, and other firing patterns.
Neural networks in the biological brain consist of a vast number of neurons, each exhibiting complex dynamical behaviors. However, when simulating large-scale neural networks, using high-dimensional models for each neuron significantly increases computational costs, limiting both simulation scale and efficiency. Therefore, adopting one-dimensional neuron models as fundamental units preserves core dynamical properties while substantially reducing computational complexity. This approach enables more efficient frameworks for brain function research, cognitive computing, and artificial neural networks, facilitating large-scale simulations. Consequently, we transform these two models into one-dimensional representations, focusing only on \( x_n \), with the corresponding equations shown in \Cref{tab:Table1}. In \Cref{Fig.2}, we analyze the bifurcation diagrams and various firing patterns of the two models. Fixing \( a = 4.1 \) and the initial value \( x_0 = 0.1 \), we present the bifurcation diagram of the Rulkov model for \( h \in [-5, 1] \) in \Cref{Fig.2}(a). The chaotic spiking patterns for \( h = -2.8 \) and \( h = 0.99 \) are illustrated in \Cref{Fig.2}(b) and (c), respectively. Similarly, with \( c = 1 \), \( b = 3 \), and \( x_0 = 0.1 \), we depict the bifurcation diagram of the Chialvo model for \( k \in [0, 1] \) in \Cref{Fig.2}(d). The resulting chaotic spiking and periodic spiking patterns for \( k = 0.01 \) and \( k = 1 \) are shown in \Cref{Fig.2}(e) and (f), respectively, demonstrating the transition between different dynamical regimes and illustrating the rich complexity of neuron firing behaviors in chaotic dynamics.

\begin{table*}[h]
\centering
\caption{Characterization of the original Chialvo and Rulkov maps with an initial condition of \( x_0 = 0.1 \).} 
\label{tab:Table1} 
\renewcommand{\arraystretch}{1.2}
\rowcolors{2}{pink!20}{yellow!15} 
\begin{tabular}{ccccc} 
\rowcolor{gray!20}
\textbf{Models} & \textbf{Equations} & \textbf{Parameters} & \textbf{Lyapunov Exponent(LE)} & \textbf{Permutation Entropy(PE)} \\
\midrule 
Chialvo & $x_{n+1}=\frac{\alpha}{1+{x_{n}}^{2}}+h$ & $a=4.1,\ h=-3.3$ & 1.09 & 1.3883 \\
Rulkov  & $x_{n+1}=cx_n^2e^{b-x_n}+k$ & $b=3,\ c=1,\ k=0.01$ & 1.01 & 1.3806 \\
\bottomrule 
\end{tabular}
\end{table*}

\subsection{Description of the MHDNN Model}
\begin{figure}[h]
\centering
\includegraphics[width=\linewidth]{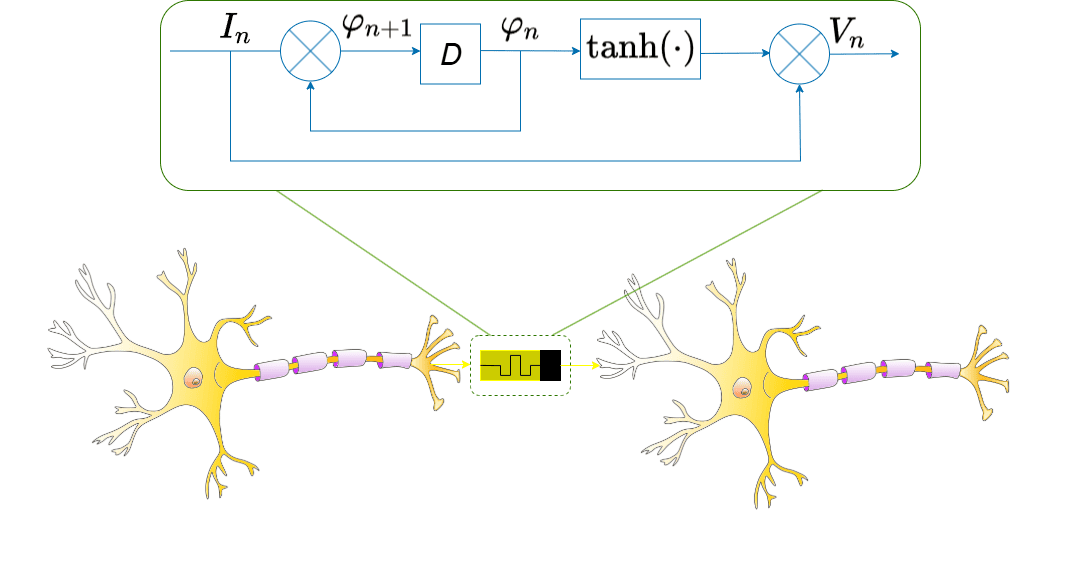}
\caption{Schematic diagram of the MHDNN.}
\label{Fig.3}
\end{figure}

The memristor, as a synaptic simulator, plays a critical role in neuromorphic computing. By dynamically adjusting its resistance, it mimics synaptic plasticity, enabling flexible information exchange between neurons. Its tunability, low power consumption, and ability for parallel processing give it significant potential to build large-scale neural networks, achieving brain-like computing, and supporting hardware-based learning. Inspired by \cite{zhou2024coexisting, xu2023initial, bao2023memristive}, we introduce  proposed discrete memristors to construct the MHDNN, with the mathematical equations given as follows:
\begin{equation}\begin{aligned}x_{n+1}&=\frac{a}{1+x_{n}^{2}}+m\tanh(z_{n})\cdot(y_{n}-x_{n})+h\\y_{n+1}&=cy_{n}^{2}\cdot e^{b-y_{n}}+m\tanh(z_{n})\cdot(y_{n}-x_{n})+3\\z_{n+1}&=z_{n}+(y_{n}-x_{n})\end{aligned}\label{2}\end{equation}
Where \( a \), \( b \), \( c \), and \( h \) are system parameters. \( x \) and \( y \) represent the membrane potentials of the Chialvo and Rulkov neurons, respectively, while \( z \) denotes the internal magnetic flux of the memristor. The parameter \( m \) controls the coupling strength between the two neurons. To clearly illustrate the structural principles of the MHDNN, its schematic diagram is shown in \Cref{Fig.3}.

\subsection{Fix point}
Fixed point analysis is a tool used to analyze dynamic systems, and the stability of the MHDNN can be determined by analyzing its fixed points. Let the fixed point set be \( P(x^*, y^*, z^*) \). From Eq. \eqref{2}, we derive:
\begin{equation}\begin{aligned}&x^{*}=\frac{a}{1+x^{*2}}+m\tanh(z^{*})\cdot(y^{*}-x^{*})+h\\&y^{*}=cy^{*2}e^{b-y^{*}}+m\tanh(z^{*})\cdot(y^{*}-x^{*})+3\\&z^{*}=z^{*}+(y^{*}-x^{*})\end{aligned}\label{3}\end{equation}\par
By solving Eq.\eqref{3}, we obtain \( P = (x^*, y^*, d) \), where \( d \) is any constant, \( x^* = y^* \), and \( x^* \) can be obtained by solving the following equation:
\begin{equation}cx^{*^{4}}e^{b-x^{*}}+cx^{*^{2}}e^{b-x^{*}}-a=0\label{4}\end{equation}\par
The corresponding Jacobian matrix can be derived from the fixed point set \( P \):
\begin{equation}J=\begin{bmatrix}J_{11}&m\tanh(d)&0\\-m\tanh(d)&J_{22}&0\\-1&1&1\end{bmatrix}\label{5}\end{equation}
Where $J_{11}=\frac{-2ax^{*}}{(1+x^{*^{2}})^{2}}-m\tanh(d)$ , $J_{22}=2cy^{*}e^{b-y^{*}}-cy^{*^{2}}e^{b-y^{*}}+m tanh(d)$. The characteristic equation can be obtained from Eq. \eqref{5}:
\begin{equation}(\lambda-1)[\lambda^2-(J_{11}+J_{22})\lambda+(J_{11}J_{22}+(m\tanh(d))^2)]=0\label{6}\end{equation}\par
The three eigenvalues can be solved as:\par
\begin{equation}\lambda_1=1\label{7}\end{equation}
\begin{equation}\lambda_2=\frac{(J_{11}+J_{22})}2-\frac{\sqrt{(J_{11}-J_{22})^2-4(m\tanh(d))^2}}2\label{8}\end{equation}
\begin{equation}\lambda_2=\frac{(J_{11}+J_{22})}2+\frac{\sqrt{(J_{11}-J_{22})^24(m\tanh(d))^2}}2\label{9}\end{equation}\par
The stability of the MHDNN system is confirmed when all three eigenvalues lie within the unit circle; otherwise, instability arises. From Eq. \eqref{7} to Eq. \eqref{9}, it is observed that \(\lambda_1\) always resides on the unit circle, while the positions of \(\lambda_2\) and \(\lambda_3\) within the unit circle depend on the parameters \(a\), \(b\), \(c\), $m$ and the initial conditions \(x\) and \(y\). Notably, \(\lambda_3 \geq \lambda_2\), allowing the region of instability to be determined as:\\
\begin{equation}\begin{aligned}&|J_{11}+J_{22}+\sqrt{(J_{11}^{2}-J_{22})^{2}+4(m\tanh(d))^{2}}|>2
\\&or
\\&|J_{11}+J_{22}-\sqrt{(J_{11}^{2}-J_{22})^{2}+4(m\tanh(d))^{2}}|>2\end{aligned}\end{equation}\par

\begin{figure}[!htp] 
\centering 
\includegraphics[width=0.5\textwidth]{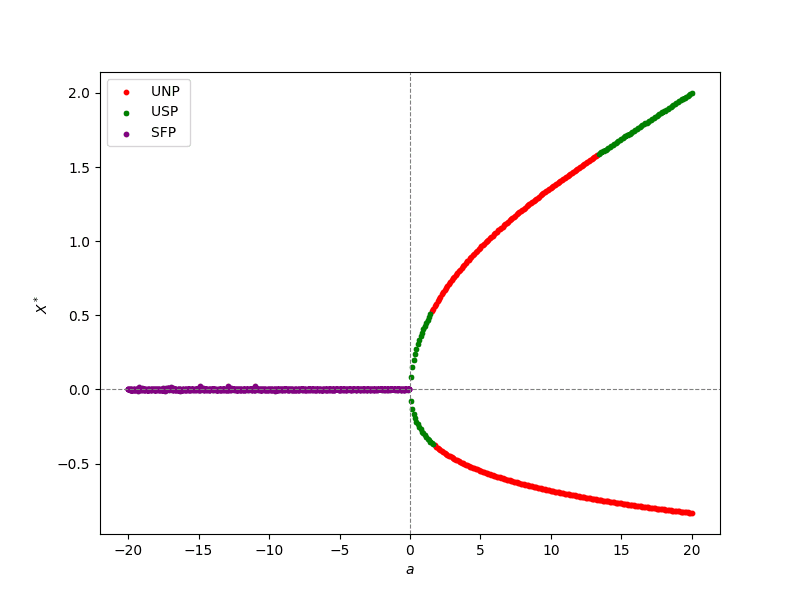} 
\caption{Different types of fixed points on the \( a - x^* \) plane.} 
\label{Fig.4} 
\end{figure}

With system parameters set as \( (b, c, m, h) = (2,1,0.2,1) \), the fixed-point trajectories distributed in the \( a \)-\( x^*\) plane are shown in \Cref{Fig.4}. The red, green, and purple points represent the unstable node point (UNP), the unstable saddle point (USP), and the stable fixed point (SFP), respectively. It can be observed that when \( a \) is negative, there are infinitely many stable points. When \( a = 0 \), a supercritical bifurcation occurs, where the stable fixed point splits into an unstable node and an unstable saddle point. In neuroscience, this phenomenon describes neuronal excitation and inhibition: when \( a < 0 \), neuronal activity remains stable in a resting state, whereas when \( a > 0 \), the neuron enters an unstable state, potentially leading to sustained oscillations (epilepsy) or activation (excitation).\\
In summary, the critical stability or instability of the MHDNN system is determined by the system parameters and initial conditions, highlighting the inherent complexity of the system.
\section{Dynamical Behavior}
\label{sec3}
\subsection{Dynamic behavior regarding parameters}
Bifurcation diagrams, LEs, and attractors are essential tools for analyzing the dynamical behavior of chaotic systems. Fixing the initial conditions at \( (x_0, y_0, z_0) = (0.1, 0.1, 0.1) \), we examine the effects of parameters \( a \) and \( m \) on system dynamics. To enhance clarity, the parameter sets are categorized into two distinct cases.\par
\begin{figure}[h]
\centering
\includegraphics[width=\linewidth]{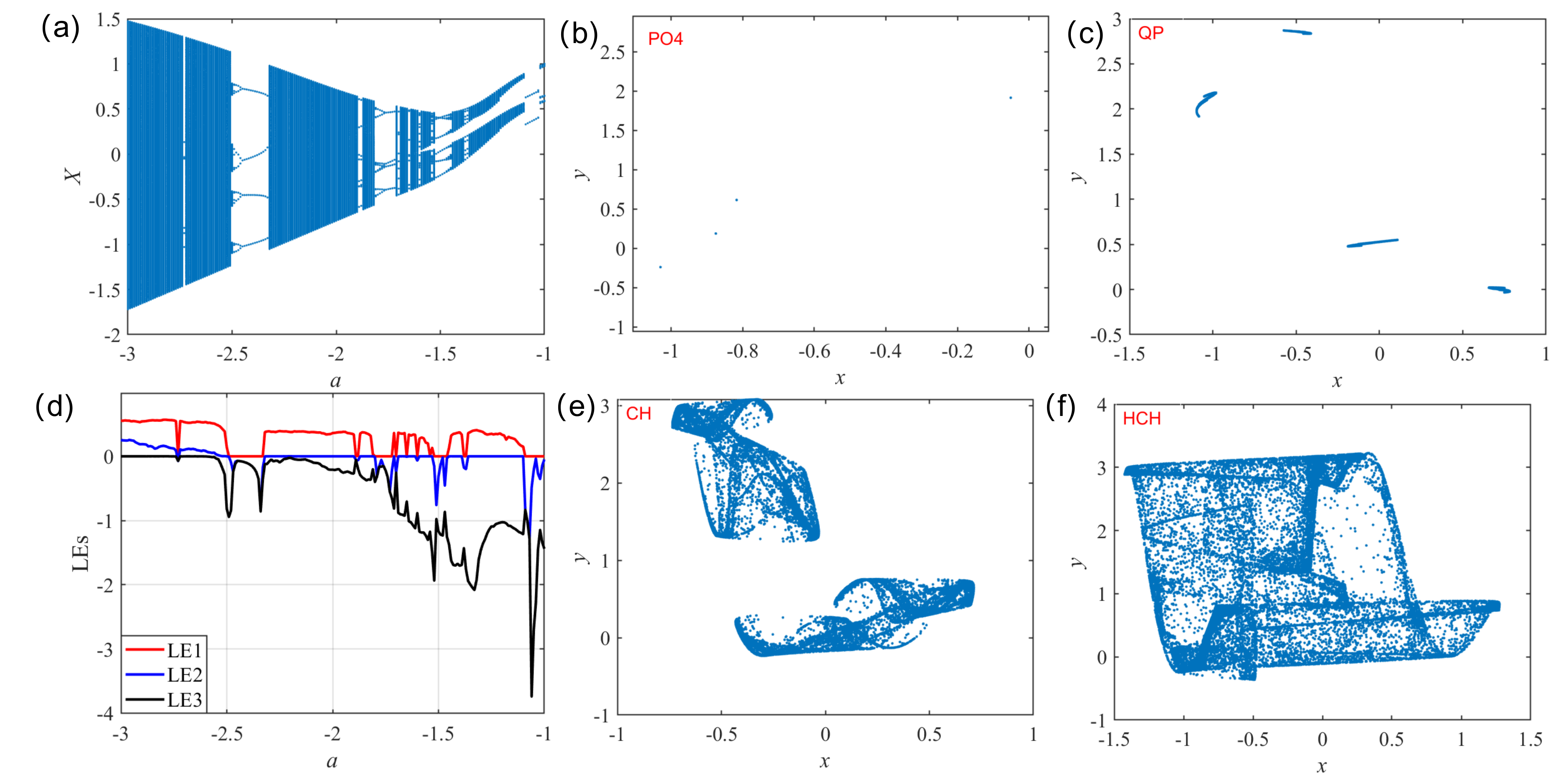}
\caption{
Bifurcation and phase diagrams with varying parameter $a$: 
(a) Bifurcation diagram; 
(b) Periodic state at $a = -2.73$; 
(c) Quasi-periodic state at $a = -2.5$; 
(d) Lyapunov exponents; 
(e) Chaotic state at $a = -2$; 
(f) Hyperchaotic state at $a = -2.7$.
}
\label{Fig.5}
\end{figure}

For Case 1, with fixed parameters \( h = 1.3 \), \( b = 1.5 \), \( c = -1.5 \), and \( m = 0.2 \), the parameter \( a \) varies within the range \( [-3, -1] \). The corresponding bifurcation diagram and LEs are presented in \Cref{Fig.5}, revealing a diverse spectrum of dynamical behaviors.
Within the interval \( a \in [-3, -1.89] \), the system predominantly displays hyperchaotic states, punctuated by the occurrence of quasi-periodic states. Additionally, in the range \( [-1.89, -1] \), chaotic and quasi-periodic states alternate, and the bifurcation diagram reveals multiple periodic windows within this range. 
As illustrated in \Cref{Fig.3}(b), (c), (e), and (f), when \( a \) is set to -2.73, -2.5, -2, and -2.7, MHDNN generates representative attractors with distinct topological structures: hyperchaotic (HCH), period-4 (P04), chaotic (CH), and quasi-periodic (QP).

\begin{figure}[h]
\centering
\includegraphics[width=\linewidth]{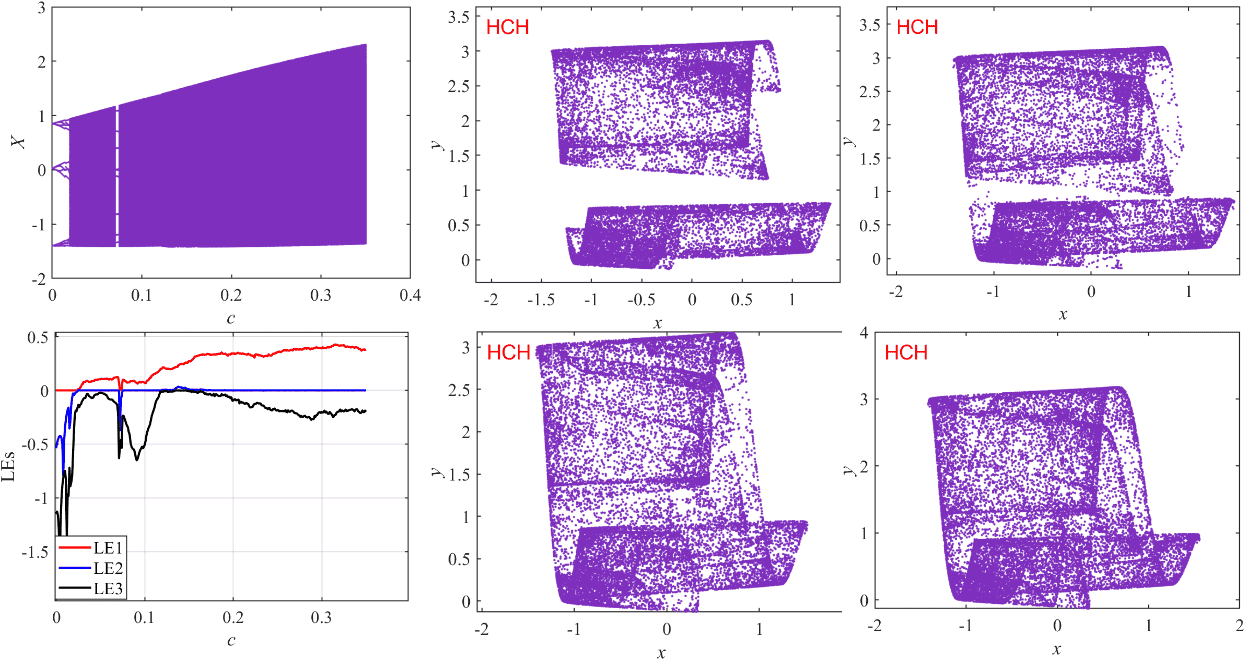}
\caption{
Bifurcation and phase diagrams with varying parameter $m$: 
(a) Bifurcation diagram; 
(b) Phase portrait at $m = 0.121$; 
(c) $m = 0.139$; 
(d) Lyapunov exponents; 
(e) $m = 0.15$; 
(f) $m = 0.159$.
}
\label{Fig.6}
\end{figure}

For Case 2, the coupling strength between neurons \( m \) varies within the range \([0, 0.35]\), while the parameters \( (a, h, b, c) = (-3.4, 2, 1.4, -1.5) \) remain fixed. The corresponding bifurcation diagram and LEs are presented in \Cref{Fig.6}(a) and (d).  
As illustrated by the attractors in \Cref{Fig.6}, as \( m \) increases, the two hyperchaotic attractors exhibit a clear tendency to merge, signifying the system's transition toward more complex dynamical behavior.\\

\subsection{Basins of attraction and firing patterns}
\begin{figure}[h]
\centering
\includegraphics[width=\linewidth]{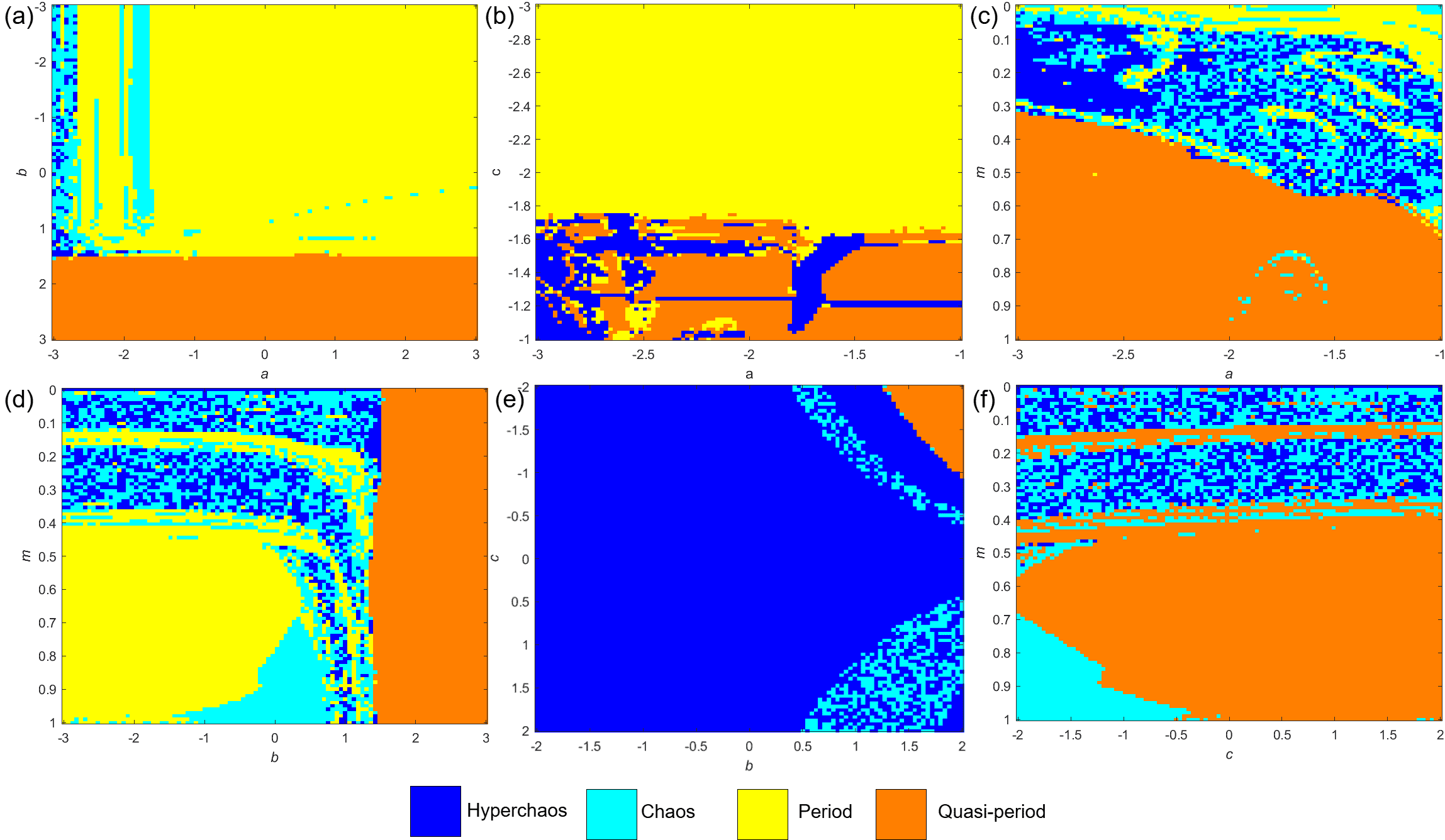}
\caption{
Basins of attraction under fixed initial condition $(x_0, y_0, z_0) = (0.1, 0.1, 0.1)$:
(a) $a$–$b$, fixed $(h, c, m) = (1.3, -1.5, 0.059)$;
(b) $a$–$c$, fixed $(h, b, m) = (2, 1.4, 0.059)$;
(c) $a$–$m$, fixed $(b, c, h) = (1.5, -1.5, 1.3)$;
(d) $b$–$m$, fixed $(a, c, h) = (-3.4, -1.5, 1.3)$;
(e) $b$–$c$, fixed $(a, h, m) = (-3.4, 1.3, 0.059)$;
(f) $c$–$m$, fixed $(a, h, b) = (-3.4, 1.3, 0)$.
}
\label{Fig.7}
\end{figure}

To further investigate how parameter variations influence the states of the MHDNN system, we fix the initial conditions at \( (x_0, y_0, z_0) = (0.1, 0.1, 0.1) \). As shown in \Cref{Fig.7}, the regions of attraction for different parameter combinations are plotted, where dark blue represents hyperchaos, light blue indicates chaos, yellow denotes periodic behavior, and orange corresponds to quasi-periodic states. By adjusting parameter combinations, a diverse range of system behaviors is revealed.\par
\begin{figure}[h]
\centering
\includegraphics[width=\linewidth]{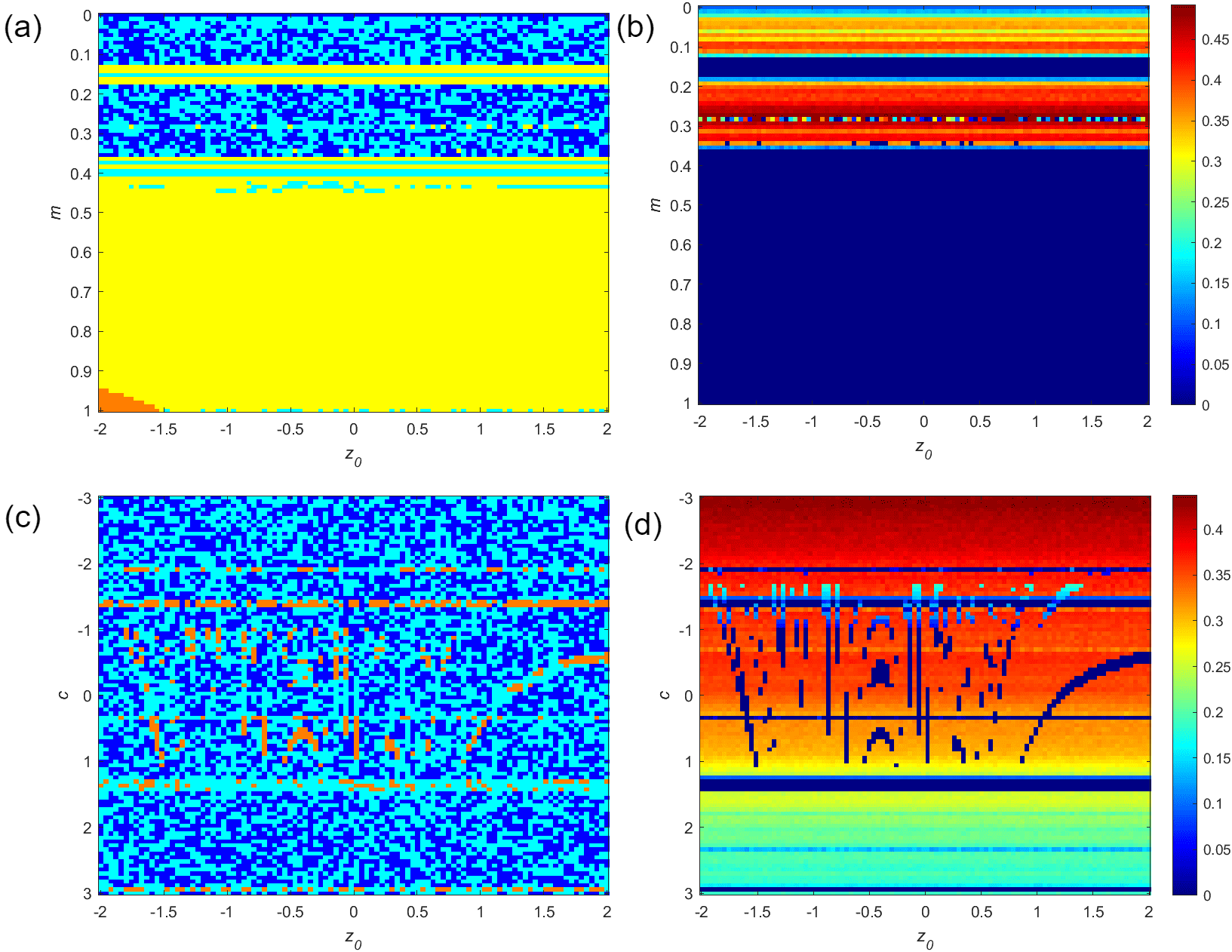}
\caption{
Memristor dynamics based on 2D bifurcation and LLE diagrams with initial state $(x_0, y_0, z_0) = (0.1, 0.1, z_0)$ and parameters $(a, b, h) = (-3.4, -1.5, 1.3)$.
(a) $z_0$–$m$ bifurcation ($c = -1.5$); 
(b) LLE diagram ($c = -1.5$); 
(c) $z_0$–$c$ bifurcation ($m = 0.34$); 
(d) LLE diagram ($m = 0.34$).
}
\label{Fig.8}
\end{figure}

Additionally, to analyze the firing patterns related to the memristor synapse, we treated \( z_0 \) and various system parameters as variables, with the initial conditions fixed at \( (x_0, y_0, z_0) = (0.1, 0.1, z_0) \) and the parameters set to \( h = 1.3 \), \( a = -3.4 \), and \( b = -1.5 \). Within the ranges \( z_0 \in [-2, 2] \), \( m \in [0, 1] \), and \( c \in [-3, 3] \), we plotted the 2D bifurcation diagrams and dynamic maps for \( z_0 - m \) and \( z_0 - c \), with the system dynamics characterized by the largest Lyapunov exponent (LLE).
As shown in \Cref{Fig.8}, different colors represent various system states. Observing the bifurcation diagrams across both planes highlights the system's complexity, irregularity, and diversity. The colored regions in the 2D LLE spectrum correspond to iterative sequences exhibiting distinct dynamic behaviors. The dark red, red, and yellow regions, indicating positive LLE, correspond to hyperchaotic or chaotic behaviors, while the blue, dark blue, and black regions, with zero LLE, correspond to periodic and quasi-periodic behaviors. These dynamics align with the patterns observed in the bifurcation diagrams.\par
\begin{figure}[h]
\centering
\includegraphics[width=\linewidth]{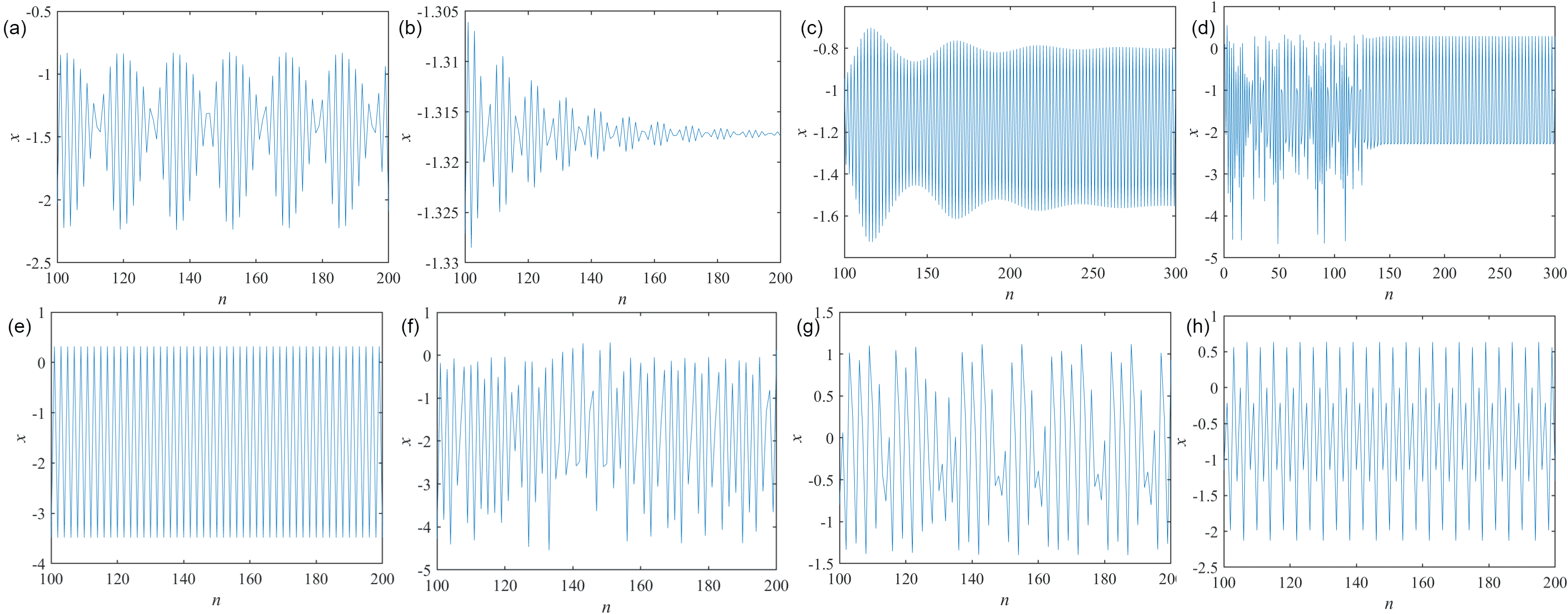}
\caption{
Firing patterns of variable $x$ in the MHDNN under different memristor parameters $(z_0, m, c)$:
(a) Hyperchaotic bursting (1, -0.3, 0.3);
(b) Post-spike hyperchaotic oscillation (1, -0.3, 0.2);
(c) Transient chaotic bursting (0.1, -0.2, 0.2);
(d) Transient quasi-periodic bursting (0.1, -0.2, 1);
(e) Quasi-periodic spiking (0.3, -0.15, 1);
(f) Chaotic spiking (0.1, -0.25, 0.8);
(g) Chaotic bursting (0.1, 0.16, 0.2);
(h) Periodic bursting (0.3, -0.1, -0.1).
}
\label{Fig.9}
\end{figure}

Furthermore, based on \Cref{Fig.8}, with the initial conditions fixed at \( (x_0, y_0, z_0) = (0.1, 0.1, z_0) \) and parameters set to \( a = -3.4 \), \( b = 3 \), and \( h = 1.3 \), we selected different parameter values to analyze eight distinct firing patterns in the MHDNN system. As shown in \Cref{Fig.9}, a variety of spiking and bursting firing patterns are observed, including quasi-periodic, periodic, chaotic, and hyperchaotic behaviors. The corresponding LEs are provided in \Cref{tab:Table2}. Thus, the MHDNN system exhibits a rich diversity of firing modes, highlighting its potential applications in neural engineering.

\begin{table}[h]
\centering
\caption{The firing patterns and LEs of eight sets of memristor parameters.}
\label{tab:Table2}
\renewcommand{\arraystretch}{1.2}
\small 
\rowcolors{2}{gray!10}{white} 
\resizebox{\columnwidth}{!}{%
\begin{tabular}{c c c}
\toprule
\textbf{($z_0$, $m$, $c$)} & \textbf{Firing patterns} & \textbf{LEs} \\
\midrule
(1, -0.3, 0.3)         & Hyperchaotic bursting            & (0.3255, 0.3255, 0) \\
(1, -0.3, 0.2)         & Post-spike hyperchaotic oscillation & (0.3239, 0.3239, 0) \\
(0.1, -0.2, 0.2)       & Transient chaotic bursting       & (0.9312, 0, -0.2664) \\
(0.1, -0.2, 1)         & Transient quasi-period bursting  & (0, -0.3154, -1.1226) \\
(0.3, -0.15, 1)        & Quasi-period spiking             & (0, -0.1197, -1.6945) \\
(0.1, -0.25, 0.8)      & Chaotic spiking                  & (0.7007, 0, -1.3964) \\
(0.1, 0.16, 0.2)       & Chaotic bursting                 & (1.2523, 0, -1.8683) \\
(0.3, -0.1, -0.1)      & Period spiking                   & (-0.0457, -0.1507, -6.4387) \\
\bottomrule
\end{tabular}
}
\end{table}

\subsection{Synchronization between two coupled neurons}
When studying the behavior of two coupled neurons, we are concerned with whether their firing patterns (time series) are synchronized, the strength of their coupling, and whether phase synchronization or lag exists after coupling. This synchronization is crucial for normal brain function and cognitive processes\cite{stowe2022diurnal}\cite{kapogiannis2011disrupted}. 
Typically, the dynamical equations of coupled neurons are nonlinear, making it challenging to directly determine their synchronization from the equations. Therefore, we require a quantitative measure, the Pearson correlation coefficient, to assess their synchrony. The calculation formula is as follows:
\begin{equation}r=\frac{\sum(x_n-\bar{x})(y_n-\bar{y})}{\sqrt{\sum(x_n-\bar{x})^2}\sqrt{\sum(y_n-\bar{y})^2}}\end{equation}
Where \( x_n \) and \( y_n \) represent the time series of the two neurons, and \( \bar{x} \) and \( \bar{y} \) are their respective means. When the two neurons are synchronized, the Pearson correlation coefficient will approach 1 or -1.
Additionally, since signal transmission between neurons is not instantaneous but involves time delays—such as the fixed time lag observed in interactions between different brain regions (e.g., from the motor cortex to the cerebellum \cite{esmaeili2021rapid})—cross-correlation is introduced to measure whether two neurons exhibit time-delay synchronization. This approach helps in understanding the information processing mechanisms of neural networks and how temporal relationships reconstruct neural circuits, revealing causal relationships in biological systems. Its mathematical expression is given by:
\begin{equation}R_{xy}^{\mathrm{norm}}(\tau)=\frac{\sum_nx(n)y(n+\tau)}{\sqrt{\sum_nx(n)^2\sum_ny(n)^2}}\end{equation}
When the absolute value of the result equals 1, the two signals are fully correlated at the time delay \( \tau \).  
The system's initial conditions are chosen as \( (0,1,0,1,0,1) \). By varying the system parameters, the two neurons can exhibit different firing patterns. The synchronization of neurons in various firing modes is analyzed. \Cref{tab:Table3} lists the firing patterns and corresponding parameters for two different cases.
\par
\begin{figure}[h]
\centering
\includegraphics[width=\linewidth]{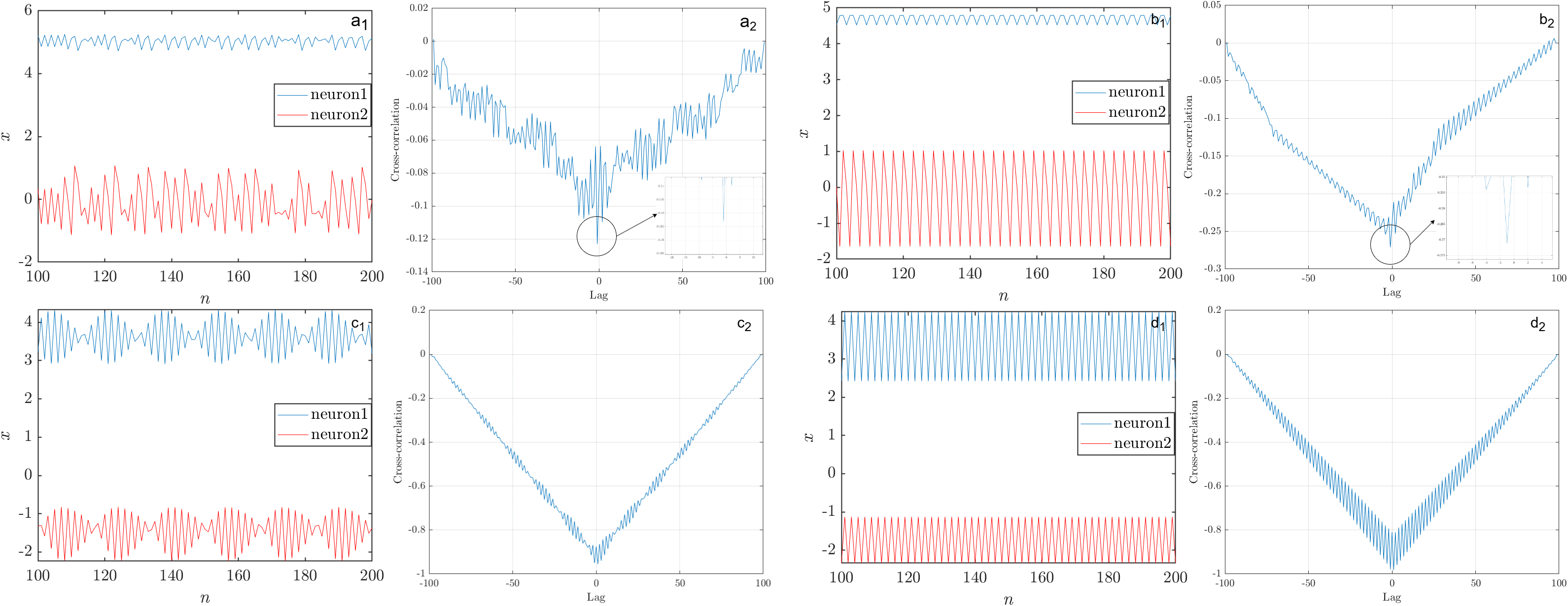}
\caption{
Evolution of membrane potential $x$ and cross-correlation in Case 1 under different coupling strengths $m$: 
(a1–a2) $m = 0.2$, 
(b1–b2) $m = 0.1$, 
(c1–c2) $m = -0.3$, 
(d1–d2) $m = -0.4$.
}
\label{Fig.10}
\end{figure}

For Case 1, a static neuron and a bursting-firing neuron are coupled through a memristor acting as a synapse. Unlike previous studies, as the coupling strength decreases, the Pearson correlation coefficient gradually approaches 1, as shown in \Cref{tab:Table4}. Furthermore, in \Cref{Fig.10}, when the coupling gain is negative, the cross-correlation reaches a higher value at negative \( \tau \), indicating that the static neuron, once activated, gradually synchronizes. This is a rare and counterintuitive phenomenon, which warrants further investigation.
\par
\begin{figure}[h]
\centering
\includegraphics[width=\linewidth]{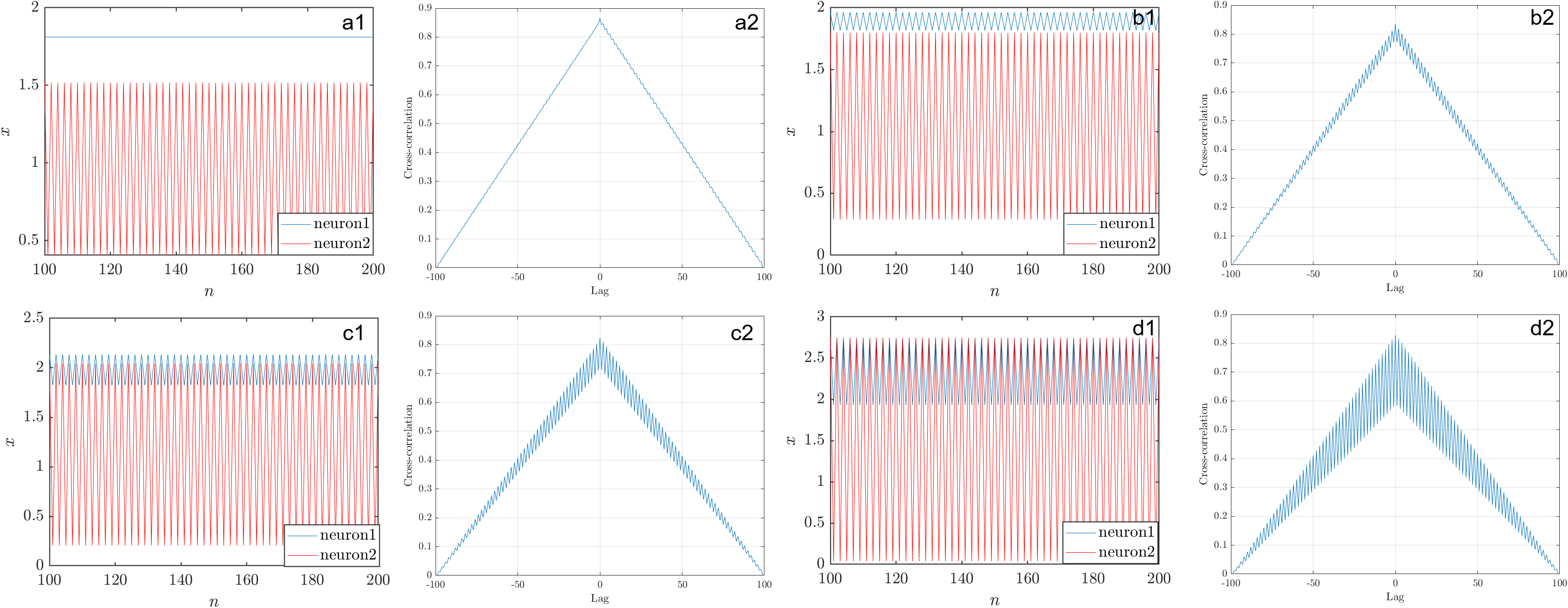}
\caption{
Membrane potential $x$ and cross-correlation of two neurons in Case 2 under different coupling strengths $m$: 
(a1–a2) $m = 0$, 
(b1–b2) $m = 0.1$, 
(c1–c2) $m = 0.2$, 
(d1–d2) $m = -0.5$.
}
\label{Fig.11}
\end{figure}

For Case 2, as shown in \Cref{Fig.11} and \Cref{tab:Table4}, synchronization cannot be achieved when the coupling strength \( m = 0 \). As \( m \) increases, the two neurons tend to synchronize, exhibiting a conventional spike-triggered pattern. Furthermore, by comparing \Cref{Fig.11}a2-d2, when the coupling strength reaches its maximum, the cross-correlation coefficient not only reflects an overall correlation trend but also reveals periodic coupling characteristics between the two signals. This suggests that neuronal interactions are not merely simple synchronous variations but are influenced by an oscillatory rhythm, a phenomenon commonly observed in neural systems such as the cortex-thalamus loop.

\begin{table}[h]
\centering
\caption{Coupling between two neurons with different firing patterns}
\label{tab:Table3}
\renewcommand{\arraystretch}{1.2}
\resizebox{\columnwidth}{!}{%
\begin{tabular}{
    >{\columncolor{gray!10}\centering\arraybackslash}c
    >{\columncolor{green!5}\centering\arraybackslash}c
    >{\columncolor{blue!5}\centering\arraybackslash}c
}
\toprule
\textbf{Case} & \textbf{Neuron 1} & \textbf{Neuron 2} \\
\midrule
\textbf{\textcolor{red}{1}} & Bursting ($a = -3.4$, $h = 1.3$) & Quiescent ($b = 3$, $c = 0.3$) \\
\textbf{\textcolor{red}{2}} & Spiking (high freq., $a = 2$, $h = -0.2$) & Fast quiescent ($b = 2$, $c = -0.3$) \\
\bottomrule
\end{tabular}%
}
\end{table}

\begin{table}[h]
    \centering
    \caption{Pearson correlation coefficients for Case 1 and Case 2}
    \label{tab:Table4}
    \renewcommand{\arraystretch}{1.1}
    \begin{tabular}{>{\columncolor{green!10}\centering\arraybackslash}p{1.2cm}
                    >{\columncolor{green!10}\centering\arraybackslash}p{2cm}
                    >{\columncolor{blue!5}\centering\arraybackslash}p{1.2cm}
                    >{\columncolor{blue!5}\centering\arraybackslash}p{2cm}}
        \toprule
        \rowcolor{gray!20}
        \multicolumn{2}{c}{\textbf{Case 1}} & \multicolumn{2}{c}{\textbf{Case 2}} \\
        \cmidrule(lr){1-2} \cmidrule(lr){3-4}
        $m$ & $r$ & $m$ & $r$ \\
        \midrule
        0.2  & 0.14214  & 0.0  & 0.06252 \\
        0.1  & -0.06170 & 0.1  & 0.77166 \\
        -0.3 & 0.74059  & 0.2  & 0.93236 \\
        -0.4 & 0.99704  & 0.5  & 0.98921 \\
        \bottomrule
    \end{tabular}
\end{table}

\section{Applications}
\label{sec4}
\subsection{PRNG based on MHDNN map}
Due to their inherent randomness and sensitivity to initial conditions, chaotic systems are widely utilized in information encryption, particularly in PRNG and image encryption. In this study, the initial conditions of the MHDNN system are set as \( (x_0, y_0, z_0) = (0.1, 0.1, 0.1) \), with parameters \( (a, h, b, m, c) = (-3.4, 1.3, 1.5, 0.2, 4.28) \), generating two hyperchaotic sequences \( X = \{X(1), X(2), \dots, X(n), \dots\} \) and \( Y = \{Y(1), Y(2), \dots, Y(n), \dots\} \).  
The sequences \( X(n) \) and \( Y(n) \) are converted into 32-bit binary streams \( X_B(n) \) and \( Y_B(n) \) using the IEEE 754 floating-point standard, each with a total length of \( 10^6 \). Pseudorandom numbers (PRNs) are then obtained by extracting bits 25 to 32 from these binary streams, mathematically expressed as:
\begin{equation}P_i(1)=X_B(n)_{25:32}\label{10}\end{equation}
\begin{equation}P_i(2)=Y_B(n)_{25:32}\label{11}\end{equation}
Next, to assess the randomness of the generated PRNs, we employed the NIST SP800-22 statistical test suite with a significance level of \( \alpha = 0.01 \). A sequence is deemed to pass the randomness test if its \( P\text{-value} > \alpha \). The results of the 15 subtests from the NIST suite are presented in \Cref{tab:Table5}. The findings indicate that the pseudorandom sequences generated by the MHDNN exhibit strong randomness, validating their applicability in image encryption.
\begin{table}[!htb]
\centering
\caption{NIST SP800-22 test results of the proposed PRNG}
\label{tab:Table5}
\rowcolors{2}{gray!10}{white} 
\resizebox{\columnwidth}{!}{
\begin{tabular}{cccccc}
\toprule
No. & Sub-tests & $P\text{-value}_x$ & Result & $P\text{-value}_y$ & Result \\
\midrule
01 & Frequency                  & 0.437274  & Pass & 0.534146  & Pass \\
02 & Block frequency           & 0.911413  & Pass & 0.834308  & Pass \\
03 & Cumulative sums           & 0.964295  & Pass & 0.911413  & Pass \\
04 & Runs                      & 0.090936  & Pass & 0.213309  & Pass \\
05 & Longest test              & 0.739918  & Pass & 0.162606  & Pass \\
06 & Rank                      & 0.834308  & Pass & 0.213309  & Pass \\
07 & FFT                       & 0.162606  & Pass & 0.637119  & Pass \\
08 & Non-overlapping template  & 0.911413  & Pass & 0.964295  & Pass \\
09 & Overlapping template      & 0.739918  & Pass & 0.275709  & Pass \\
10 & Universal                 & 0.834308  & Pass & 0.637119  & Pass \\
11 & Approximate entropy       & 0.739918  & Pass & 0.637119  & Pass \\
12 & Random excursions         & 0.350485  & Pass & 0.437274  & Pass \\
13 & Random excursion variant  & 0.213309  & Pass & 0.275709  & Pass \\
14 & Serial                    & 0.911413  & Pass & 0.213309  & Pass \\
15 & Linear complexity         & 0.437274  & Pass & 0.350485  & Pass \\
\bottomrule
\end{tabular}
}
\end{table}

\subsection{Hardware implementation}
\begin{figure}[!htp] 
\centering 
\includegraphics[width=0.5\textwidth]{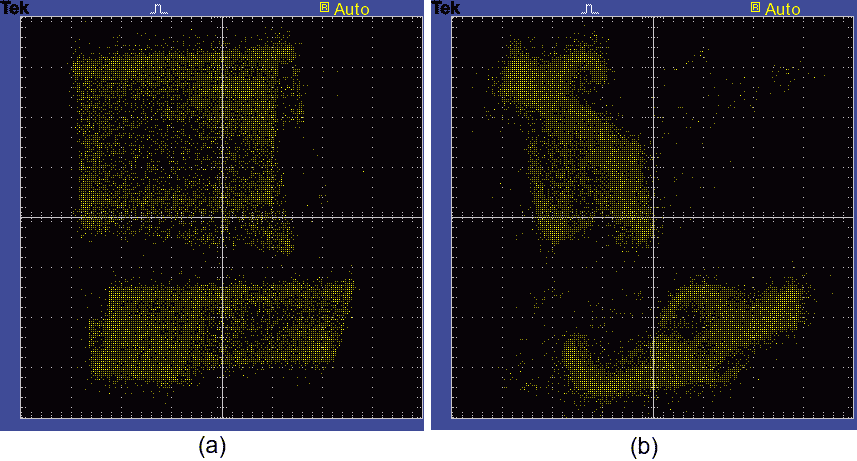} 
\caption{Phase portrait plots of hyperchaotic attractors acquired from the digital hardware prototype under different parameters. (a) (a, h, b, c, m) = (-3.4, 2, 1.4, -1.5, 0.121). (b) (a, h, b, c, m) = (-2, 1.3, 1.5, -1.5, 0.2).} 
\label{Fig.12} 
\end{figure}

Digital circuits offer advantages over analog circuits, such as enhanced noise immunity, repeatability, programming flexibility, and ease of storage and transmission. Therefore, in this section, we implement the MHDNN on the STM32 platform, specifically using the STM32F407VET6 chip and the AD5689 D/A converter. As shown in \Cref{Fig.12}, the attractor captured by the oscilloscope corresponds to the one displayed in MATLAB.
\subsection{A novel image encryption algorithm based on the MHDNN}
In this section, we present our encryption algorithm, which is primarily based on MHDNN, S-box transformations, diffusion, and permutation operations. The detailed steps are as follows:\par

\textbf{Chaotic Keystream Generation (CKG).}%
\label{subsec:ckg-description}
The CKG module initializes the two–dimensional Lorenz–inspired recurrent neural map (MHDNN) with the secret key~$K$, discards its transient behaviour, and continuously samples the orbit to build a high-entropy chaotic sequence.  
Through deterministic scaling and modular operations, this sequence is transformed into three pixel–permutation vectors $(X,\,X_1,\,X_2)$ and a diffusion keystream $SBT$, which jointly serve as the cryptographic key material for subsequent stages.  
The complete procedure is presented in Algorithm~\Cref{Algorithm:1}.

\begin{algorithm}[H]
\caption{Chaotic Keystream Generation (CKG)}
\label{Algorithm:1}
\begin{algorithmic}[1]           
  \Require Secret key $K$; image size $M\times N$
  \Ensure  Permutation sequences $X,X_1,X_2$ and diffusion keystream $SBT$
  \State Initialize the MHDNN system with key $K$
  \For{$i \gets 1$ \textbf{to} $1000$}
        \State Iterate MHDNN once and discard the output                 \Comment{remove transient}
  \EndFor
  \For{$i \gets 1$ \textbf{to} $3MN + 512$}
        \State $s(i) \gets$ next MHDNN output                            \Comment{collect chaos}
  \EndFor
  \For{$i \gets 1$ \textbf{to} $MN$}
        \State $X(i)   \gets \operatorname{mod}\!\bigl(\lfloor(s_i+100)10^{10}\rfloor , MN\bigr) + 1$
        \State $X_1(i) \gets \operatorname{mod}\!\bigl(\lfloor(s_{i+MN}+100)10^{10}\rfloor , MN\bigr) + 1$
        \State $X_2(i) \gets \operatorname{mod}\!\bigl(\lfloor(s_{i+2MN}+100)10^{10}\rfloor , MN\bigr) + 1$
        \State $SBT(i)\gets \operatorname{mod}\!\bigl(\lfloor(s_{i+3MN}+100)10^{10}\rfloor , 256\bigr) + 1$
  \EndFor
  \State \Return $X, X_1, X_2, SBT$
\end{algorithmic}
\end{algorithm}
\textbf{Dynamic S-Box Construction (DSB).}%
To reinforce the non-linearity of the cipher, a data-dependent $16\times16$ substitution box is generated on-the-fly.  
Specifically, a fresh chaotic subsequence is partitioned, sorted to obtain index vectors, and reshaped into an intermediate matrix that is further permuted by double–swap operations.  
The resulting lookup table~$S$ exhibits good statistical properties and thwarts differential attacks.  
The step-by-step routine is given in\Cref{Algorithm2}.

\begin{algorithm}[H]
\caption{Dynamic S-Box Construction (DSB)}
\label{Algorithm2}
\begin{algorithmic}[1]
  \Require Chaotic sequence $q$ (first $3000$ iterations already discarded)
  \Ensure  $S$ — final $16 \times 16$ S-Box
  \State $\textit{list}  \gets q(1{:}256)$
  \State $\textit{list}_1\gets q(257{:}384)$
  \State $\textit{list}_2\gets q(385{:}512)$
  \State $[L,\;\_]   \gets \text{sort}(\textit{list})$
  \State $[L_1,\;\_] \gets \text{sort}(\textit{list}_1)$
  \State $[L_2,\;\_] \gets \text{sort}(\textit{list}_2)$
  \State $S_1 \gets \text{reshape}(L,16,16)^\top$
  \State $s_1 \gets \text{vec}(S_1)$
  \For{$i \gets 1$ \textbf{to} $128$}
        \State swap$(s_1,L_1(i),L_2(i))$
  \EndFor
  \State $S \gets \text{reshape}(s_1,16,16)$
  \State \Return $S$
\end{algorithmic}
\end{algorithm}

\textbf{Color Image Encryption (CIE).}
With the permutation vectors, diffusion keystream, and dynamic S-Box at hand, the CIE stage encrypts each colour channel independently.  
The plaintext vector is first scrambled by forward permutation, followed by bidirectional diffusion in which the S-Box output and keystream are XOR-combined with the running cipher state.  
A reverse permutation finalises the confusion process before the channel is reshaped back to its two-dimensional form.  
The operations for all three channels collectively constitute the overall cipher, as detailed in Algorithm~\ref{Algorithm.3}.
\begin{algorithm}[H]
\caption{Color Image Encryption (CIE)}
\label{Algorithm.3}
\begin{algorithmic}[1]
  \Require Plain image $P \in \mathbb{Z}_{256}^{M \times N \times 3}$
  \Statex \hspace{\algorithmicindent}Keystreams $X, X_1, X_2, SBT$
  \Statex \hspace{\algorithmicindent}S-Box $S$
  \Ensure  Cipher image $C$

  \State $sx \gets \text{vec}(S)$ 
  \For{\textbf{each} channel $ch \in \{R,G,B\}$}
        \State $A \gets \text{vec}(P(:,:,ch))$; \quad $T \gets MN$
        
        \For{$i \gets 1$ \textbf{to} $\lfloor T/2 \rfloor$}
              \State swap$(A,X(i),X(T-i+1))$
        \EndFor
       
        \State $B(1) \gets A(1)$
        \For{$i \gets 2$ \textbf{to} $T$}
              \State $B(i) \gets A(i) \oplus \bigl(B(i-1) \oplus sx(SBT(i))\bigr)$
        \EndFor
      
        \State $C_{\text{tmp}}(T) \gets B(T)$
        \For{$i \gets T-1$ \textbf{downto} $1$}
              \State $C_{\text{tmp}}(i) \gets B(i) \oplus \bigl(C_{\text{tmp}}(i+1) \oplus sx(SBT(i))\bigr)$
        \EndFor
        
        \For{$i \gets 1$ \textbf{to} $\lfloor T/2 \rfloor$}
              \State swap$(C_{\text{tmp}}, X_2(i), X_2(T-i+1))$
        \EndFor
        \State $C(:,:,ch) \gets \text{reshape}(C_{\text{tmp}}, M, N)$
  \EndFor
  \State \Return $C$
\end{algorithmic}
\end{algorithm}

\subsection{Simulation results}
In this section, various methods were employed to comprehensively evaluate the security of the encryption algorithm, including key space analysis, histogram analysis, correlation analysis, and information entropy, among others. All experiments were conducted on a Windows 11 system with a 12th Gen Intel(R) Core(TM) i7-12700H 2.30 GHz processor, 16GB of memory, and Matlab R2021b. The encryption process consistently used the key \( (x_0, y_0, z_0, a, b, c, h, m) = (0.1, 0.1, 0.1, -3.4, -1.5, -1.5, 3, 0.5) \).
\subsubsection{visual effect}
\begin{figure}[h]
\centering
\includegraphics[width=\linewidth]{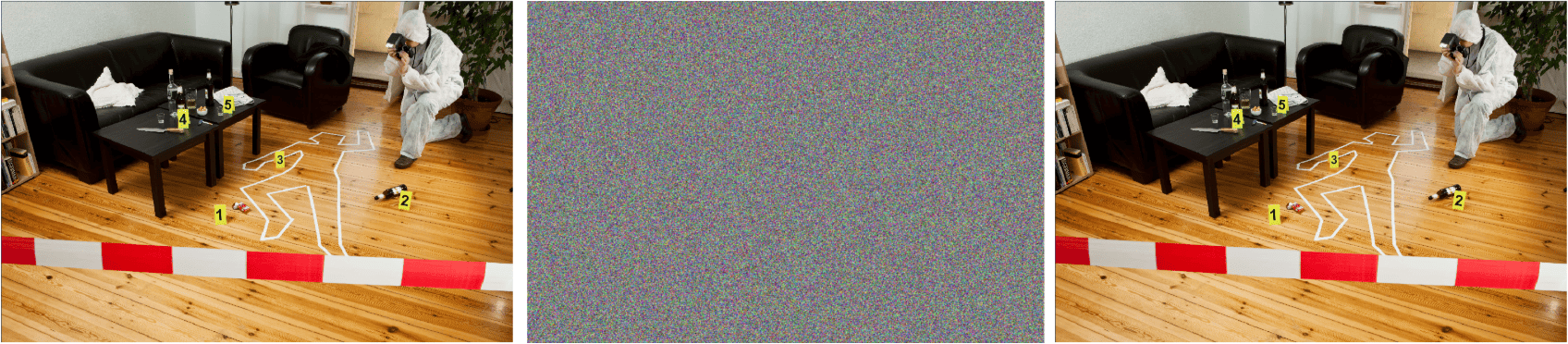}
\caption{
Image encryption and decryption results: 
(a) Original image; 
(b) Encrypted image; 
(c) Decrypted image.
}
\label{Fig.13}
\end{figure}

As shown in \Cref{Fig.13}, our proposed encryption algorithm effectively disrupts the image information, and with the correct key, it accurately restores the image. This demonstrates both the security and reversibility of the proposed encryption algorithm.

\subsubsection{Key spaces}
The key space refers to the set of all possible keys in an encryption algorithm. Each key corresponds to a specific method of encryption or decryption, and the size of the key space is typically determined by the length of the key and the possible values for each key. A larger key space enhances security by significantly increasing the number of keys an attacker must try, making exhaustive searching within a reasonable timeframe virtually impossible using modern computational resources. 
In this study, we set the maximum precision of the keys \( x_0, y_0, z_0, a, b, c, h, m \) to \( r = 16 \), resulting in a key space of \( 10^{128} \), which far exceeds \( 2^{100} \). Therefore, the proposed encryption algorithm provides robust security and is highly resistant to brute-force attacks.
\subsubsection{Histogram analysis}
\begin{figure}[!hbtp] 
\centering 
\includegraphics[width=0.5\textwidth]{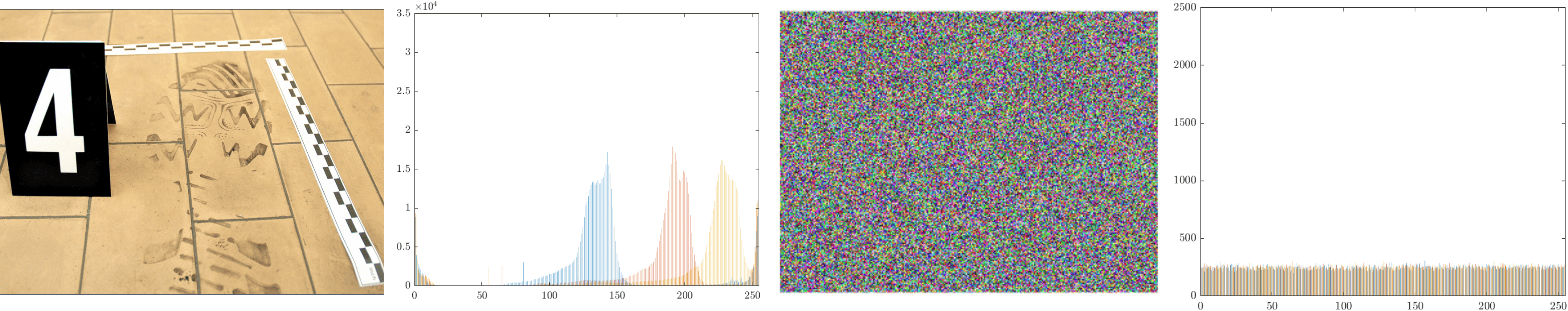} 
\caption{Histogram analysis results. (a) Original image; (b) Histogram of the original image; (c) Encrypted image; (d) Histogram of the encrypted image.} 
\label{Fig.14} 
\end{figure}

\begin{figure}[!hbtp] 
\centering 
\includegraphics[width=0.5\textwidth]{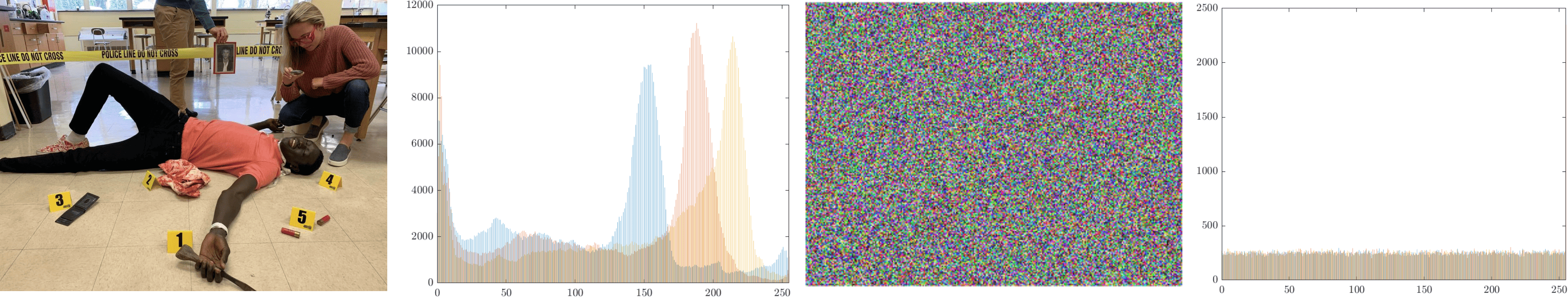} 
\caption{Histogram analysis results. (a) Original image; (b) Histogram of the original image; (c) Encrypted image; (d) Histogram of the encrypted image.} 
\label{Fig.15} 
\end{figure}

\begin{figure}[!hbtp] 
\centering 
\includegraphics[width=0.5\textwidth]{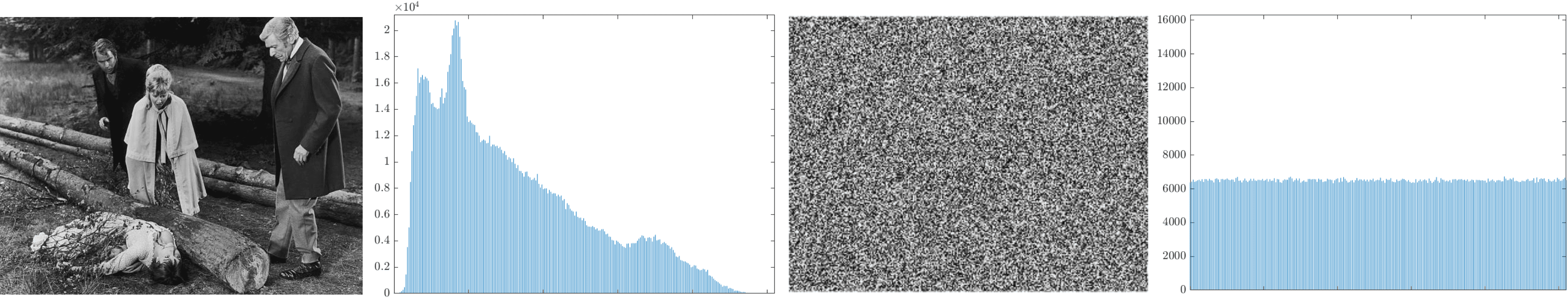} 
\caption{Histogram analysis results. (a) Original image; (b) Histogram of the original image; (c) Encrypted image; (d) Histogram of the encrypted image.} 
\label{Fig.16} 
\end{figure}

The histogram of an image reflects its pixel distribution. For the original image, the histogram distribution is typically uneven and exhibits clustering phenomena. However, if the histogram distribution becomes uniform and smooth after encryption, it indicates the effectiveness of the encryption algorithm. 
As shown in \Cref{Fig.14}-\Cref{Fig.16}, the histogram after encryption is highly uniform and free from discernible statistical patterns. This demonstrates that the encryption algorithm is effective in resisting statistical attacks.
\subsubsection{Analysis of correlation}
\begin{figure}[h]
\centering
\includegraphics[width=\linewidth]{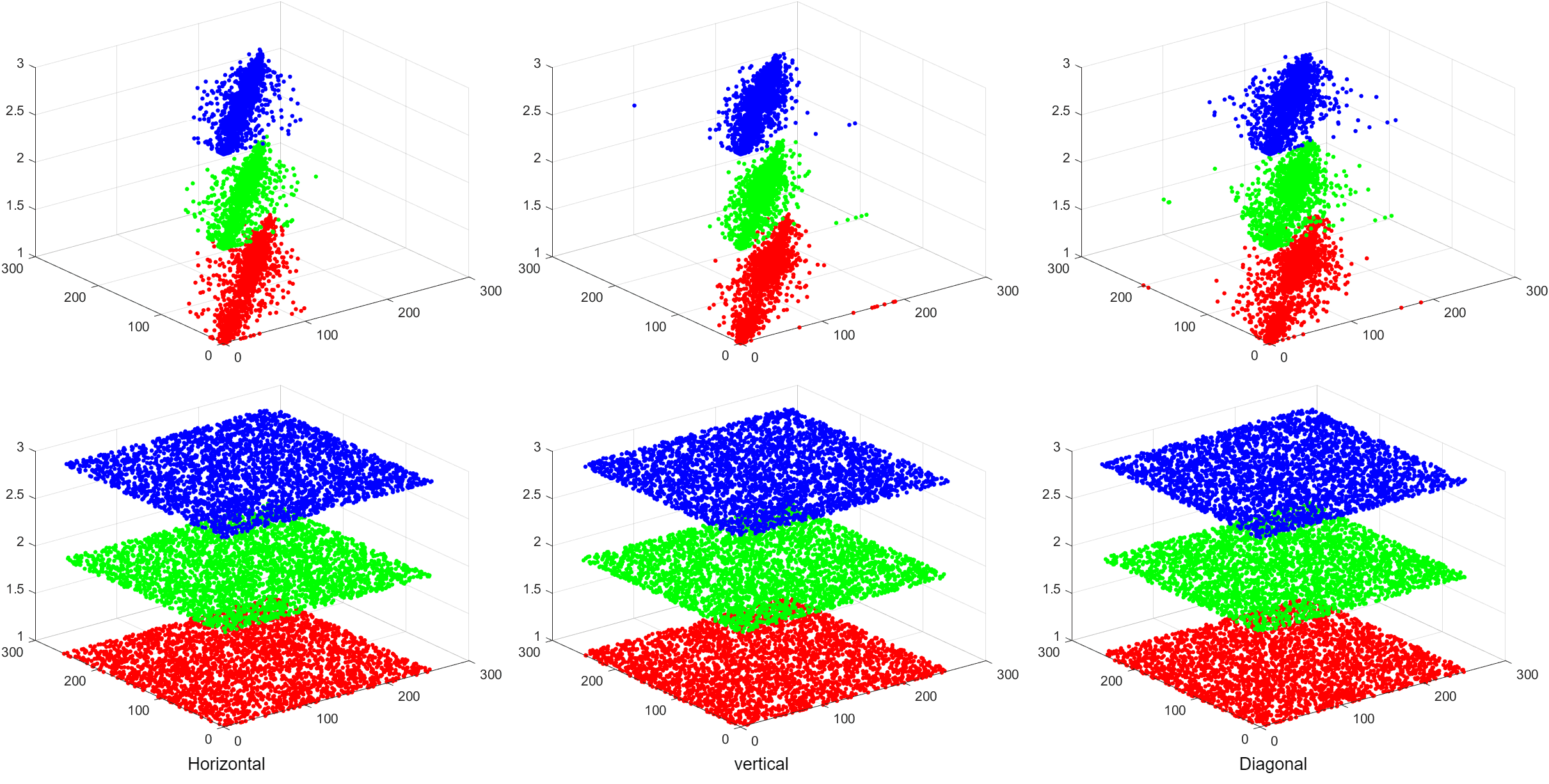}
\caption{Correlation analysis in different directions.}
\label{Fig.17}
\end{figure}

An excellent encryption algorithm should disrupt the adjacent elements of the original image, thereby reducing its correlation. The formula for calculating the correlation coefficient is given in Eq. \eqref{23}. We selected the Lena image for the experiment and tested the corresponding correlation coefficients in the horizontal, vertical, and diagonal directions. The test results, shown in \Cref{Fig.17}, indicate that the correlation of the original image is significantly reduced after applying the encryption algorithm. 
Additionally, as presented in \Cref{Table6}, we compared our method with other encryption algorithms, demonstrating that our proposed encryption algorithm effectively reduces the correlation coefficient, highlighting its robustness in resisting correlation-based attacks.
\begin{align}
r_{xy} &= \frac{|\mathrm{cov}(x,y)|}{\sqrt{D(x)} \times \sqrt{D(y)}} \nonumber \\
\mathrm{cov}(x,y) &= \frac{1}{N} \sum_{i=1}^N (x_i - E(x))(y_i - E(y)) \nonumber \\
E(x) &= \frac{1}{N} \sum_{i=1}^N x_i \nonumber \\
D(x) &= \frac{1}{N} \sum_{i=1}^N \left(x_i - E(x)\right)^2
\label{23}
\end{align}

\begin{table}[!hbtp]
\caption{Comparison with Other Encryption Algorithms}
\label{Table6}
\centering
\renewcommand{\arraystretch}{1.2}
\small 
\resizebox{\columnwidth}{!}{%
\begin{tabular}{>{\centering\arraybackslash}p{3.2cm}
                >{\centering\arraybackslash}p{2.2cm}
                >{\centering\arraybackslash}p{2.2cm}
                >{\centering\arraybackslash}p{2.2cm}}
\toprule
\rowcolor{gray!20}
\textbf{Algorithms} & \multicolumn{3}{c}{\textbf{Lena}} \\
\rowcolor{gray!20}
& \textbf{Horizontal} & \textbf{Vertical} & \textbf{Diagonal} \\
\midrule
\rowcolor{green!10}
Proposed & 0.0002 & 0.0061 & -0.0023 \\
\rowcolor{blue!5}
\textcolor{blue}{Ref.~\cite{ye2023asymmetric}} & -0.0099 & 0.0101 & 0.0037 \\
\rowcolor{blue!5}
\textcolor{blue}{Ref.~\cite{liu2016encryption}} & -0.0013 & 0.0018 & 0.0023 \\
\rowcolor{blue!5}
\textcolor{blue}{Ref.~\cite{chen2024image}} & 0.0017 & -0.0032 & 0.0021 \\
\bottomrule
\end{tabular}
}
\end{table}

\subsubsection{Information entropy}
Information entropy is a crucial metric for evaluating the effectiveness and security of image encryption algorithms. In practical applications, combining information entropy with other security indicators allows for a more comprehensive assessment of the reliability of image encryption. The formula for calculating information entropy is as follows:

\begin{equation}H(C)=-\sum_{i=0}^{255}C(i)\log_2C(i)\end{equation}
where \( C \) is an image containing \( M \times N \) pixels, and the probability distribution is given by \( C(i) = \frac{N_i}{M \cdot N} \), representing the probability of occurrence of the \( i \)-th pixel. The closer the value of information entropy is to 8, the higher the randomness of the image.
As shown in \Cref{tab:Table7}, we selected three images for entropy testing experiments and compared them with other encryption algorithms. The results indicate that the information entropy of the images encrypted using our proposed algorithm is closer to 8, demonstrating the security and effectiveness of our approach. This suggests that our encryption method successfully increases the randomness of the image, making it more resistant to cryptanalysis.
\begin{table}[!hbtp]
\centering
\caption{Information entropy comparison} 
\label{tab:Table7} 
\renewcommand{\arraystretch}{1.25} 
\small 
\begin{tabular}{>{\columncolor{gray!20}}c>{\columncolor{gray!20}\centering\arraybackslash}c>{\columncolor{gray!20}\centering\arraybackslash}c} 
\toprule 
\textbf{Image} & \textbf{Plaintext} & \textbf{Proposed} \\ 
\midrule
\rowcolor{blue!5}
Lena & 5.6822 & 7.9998 \\ 
\rowcolor{blue!5}
Cablecar & 7.4237 & 7.9997 \\ 
\rowcolor{blue!5}
BoatsColor & 7.2913 & 7.9999 \\ 
\rowcolor{blue!5}
Mean & -- & 7.9998 \\ 
\rowcolor{green!10}
\cite{liu2016encryption} & -- & 7.9983 \\ 
\rowcolor{green!10}
\cite{wang2018novel} & -- & 7.9892 \\ 
\rowcolor{green!10}
\cite{liang2022image} & -- & 7.9985 \\ 
\bottomrule
\end{tabular}
\end{table}

\subsubsection{Analysis of anti-differential attack}
Differential attacks are a powerful cryptographic analysis tool used to assess and exploit the security of cryptographic algorithms. The effectiveness of such attacks relies on identifying patterns between specific input differences and their corresponding output differences. By randomly altering a single pixel in the original image and encrypting both plaintexts with the encryption algorithm, two corresponding ciphertexts are generated. The differences between these ciphertexts can then be analyzed to infer the key \cite{kong2024memristor}.
Two key metrics, Non-Plaintext Change Rate (NPCR) and Unified Average Changing Intensity (UACI), are commonly used to evaluate the security of the encryption algorithm. These metrics measure the algorithm's resistance to attacks by examining how small changes in the plaintext influence the ciphertext. The corresponding formulas are as follows:
\begin{equation}NPCR=\frac{\sum_i^M\sum_j^NW(i,j)}{M\times N}\end{equation}
\begin{equation}UACI=\frac{\sum_i^M\sum_j^N(C(i,j)-C'(i,j))/255}{M\times N}\end{equation}
Where \( C(i,j) \) and \( C'(i,j) \) represent two encrypted images that differ by only one pixel. and
\begin{equation}W(i,j)=\begin{cases}0& C_1(i,j)=C_2(i,j);\\1&\text{otherwise.}\end{cases}\end{equation}

As shown in \Cref{tab:Table8}, we selected Lena (512×512) and BoatsColor (780×576) as the subjects for our experiments. Compared to other algorithms, the NPCR and UACI values of the encrypted images generated by our proposed encryption algorithm are closer to the ideal values of 0.996094 and 0.334635, respectively. This indicates that our proposed algorithm effectively resists differential attacks.
\begin{table}[!hbtp]
\centering
\caption{NPCR and UACI}
\label{tab:Table8}
\renewcommand{\arraystretch}{1.2}
\small 
\begin{tabular}{>{\columncolor{gray!20}}c>{\columncolor{gray!20}\centering\arraybackslash}c>{\columncolor{gray!20}\centering\arraybackslash}c}
\toprule
\textbf{Image} & \textbf{NPCR} & \textbf{UACI} \\
\midrule
\rowcolor{pink!20}
Lena & 0.996100 & 0.334871 \\
\rowcolor{pink!20}
BoatsColor & 0.996042 & 0.334895 \\
\rowcolor{green!15}
\textcolor{blue}{Ref.~\cite{sun2024dynamical}} & 0.995900 & 0.334800 \\
\rowcolor{green!15}
\textcolor{blue}{Ref.~\cite{teng2021color}} & 0.996235 & 0.334620 \\
\rowcolor{green!15}
\textcolor{blue}{Ref.~\cite{liang2022image}} & 0.996078 & 0.334875 \\
\bottomrule
\end{tabular}
\end{table}

\subsubsection{Analysis of robustness}
\begin{figure}[!hbtp] 
\centering 
\includegraphics[width=0.5\textwidth]{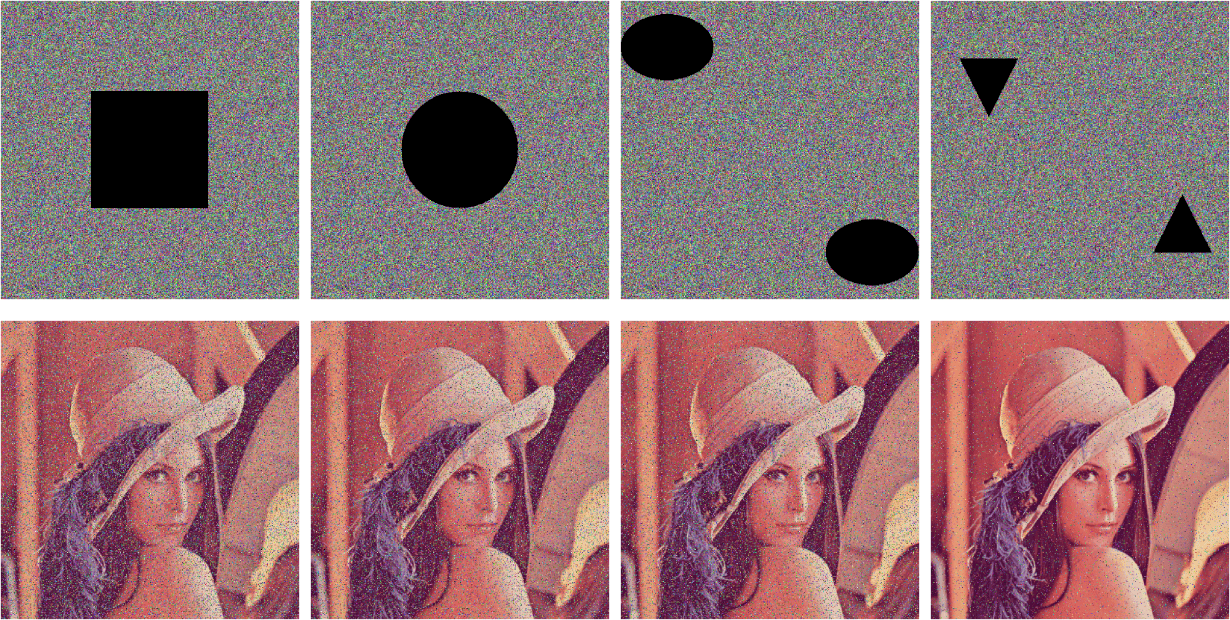} 
\vspace{-4mm}
\caption{Cropping of encrypted images with different shapes and sizes and their corresponding decrypted images.} 
\label{Fig.18} 
\vspace{-4mm}
\end{figure}

\begin{figure}[!hbtp] 
\centering 
\includegraphics[width=0.5\textwidth]{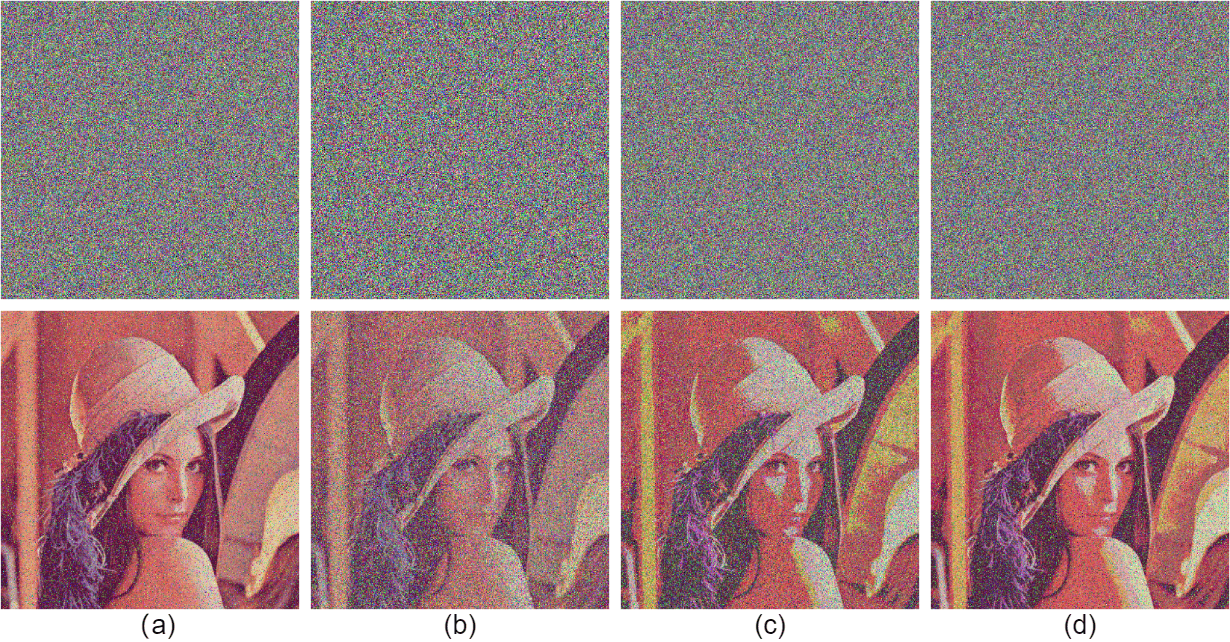} 
\vspace{-4mm}
\caption{(a) 10\% SPN; (b) 30\% SPN; (c) GN (mean 0, variance 0.01); (d) SN (variance 0.01).} 
\label{Fig.19} 
\vspace{-4mm}
\end{figure}

In  law enforcement applications, it is essential to ensure that images can be accurately recovered even when affected by noise or obstructions during transmission. To evaluate the robustness of our proposed encryption algorithm, we subjected the Lena image to various noise attacks, including salt-and-pepper noise (SPN), Gaussian noise (GN), speckle noise (SN), and different cropping attacks, as shown in \Cref{Fig.18} and \Cref{Fig.19}. The results demonstrate that the decrypted images remain visually recognizable, validating the robustness of our proposed algorithm.

\subsubsection{Image encryption application implementation}
With the advent of the era of big data, encrypting police data has become increasingly critical \cite{yu2024multiscroll}. First, it safeguards sensitive information, ensuring that only authorized personnel can access key data. This is essential for maintaining the integrity of ongoing investigations and protecting the privacy of victims and witnesses. Second, encryption prevents data breaches, reducing the risk of misuse by malicious actors such as hackers or criminals. By securing police data, law enforcement agencies can enhance operational efficiency and build public trust. Finally, data encryption ensures compliance with legal and regulatory requirements, mitigating the risk of legal liability from data loss or breaches. \par
\begin{figure*}[!hbtp] 
\centering 
\includegraphics[width=1\textwidth]{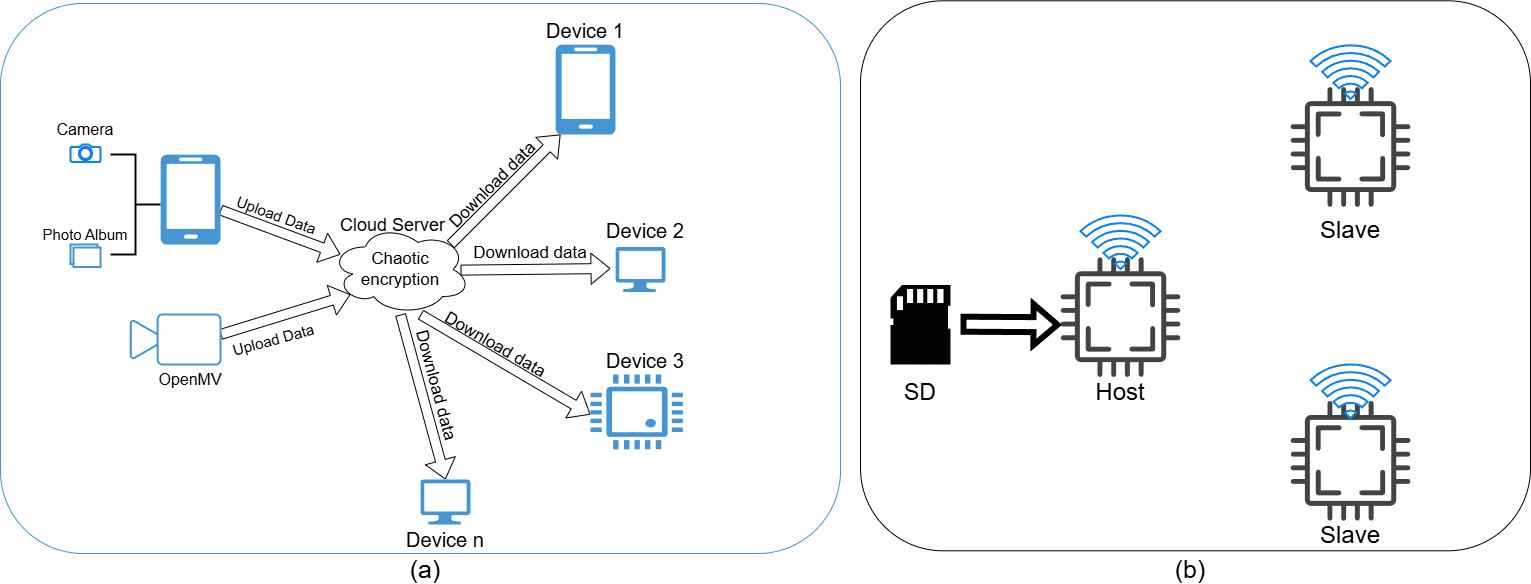} 
\caption{The process of encrypting police data. (a) Cloud encryption; (b) WiFi real-time encryption.} 
\label{Fig.20} 
\end{figure*}
\begin{figure}[!hbtp] 
\centering 
\includegraphics[width=0.5\textwidth]{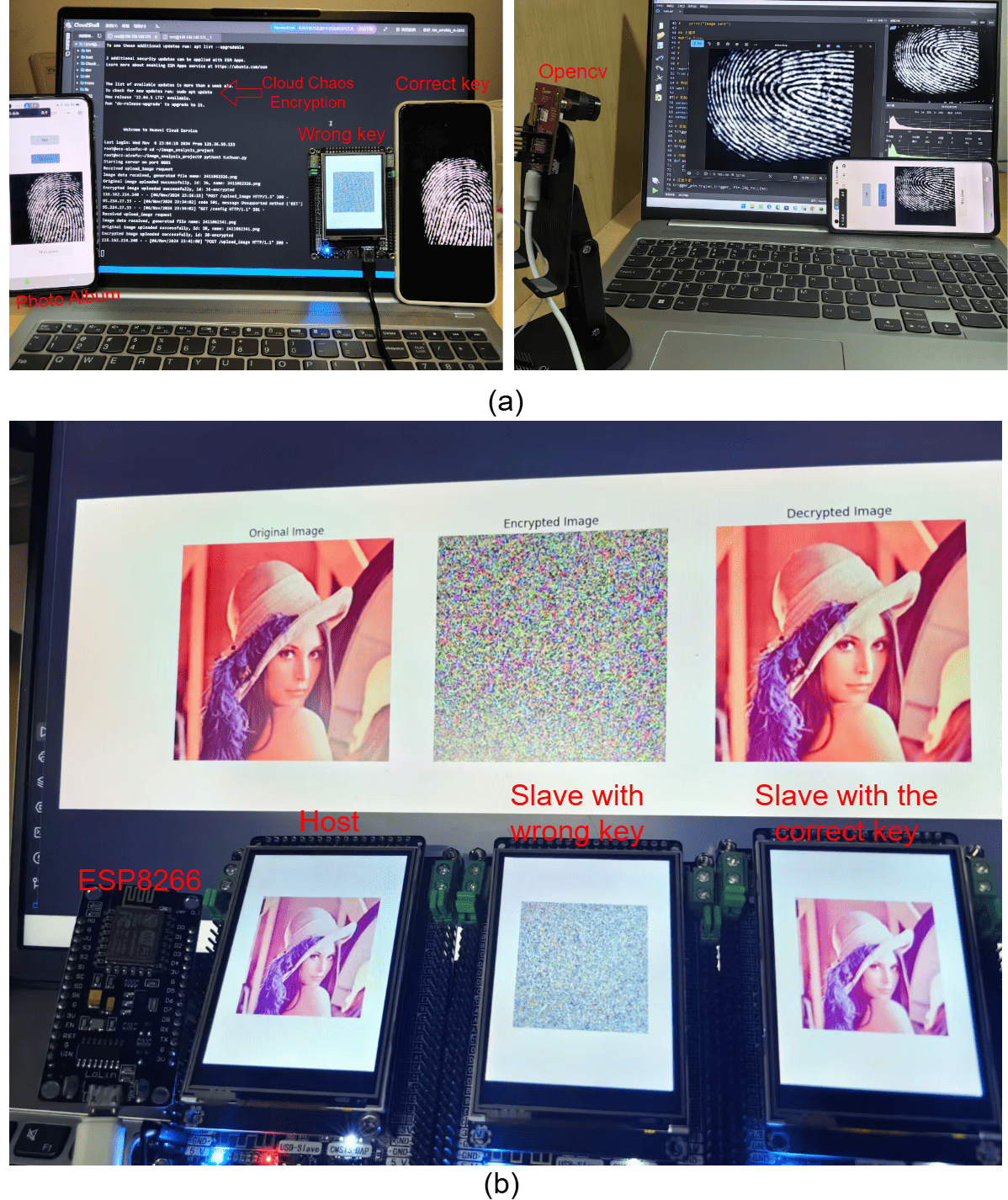} 
\caption{Hardware device used to encrypt police data. (a) Cloud-encrypted hardware device; (b) WiFi real-time encryption hardware device.} 
\label{Fig.21} 
\end{figure}

Against this backdrop, we propose a flexible police image encryption platform for the Police Internet of Things (PIoT) system based on the STM32F407VET6 microcontroller. This platform enables law enforcement officers to adapt to different operational scenarios by utilizing various devices for secure image transmission. It supports real-time image capture and encryption via a mobile camera and OpenMV, as well as encryption of pre-stored images from an SD card for later use, ensuring adaptability in diverse field conditions. The STM32F407 series microcontroller features an ARM Cortex-M4 core with a maximum clock speed of 168 MHz. Its peripherals include three 2.8-inch TFT touchscreens, an OpenMV camera module, and an ESP8266 WiFi module. The implementation flowcharts for the two methods are illustrated in \Cref{Fig.20}. Below, we provide a detailed introduction to each method: 

Case 1: Cloud Encryption
This approach utilizes smartphones, OpenMV devices, and cloud servers for secure image encryption, as illustrated in \Cref{Fig.20}(a). The server, configured with the IP address 139.159.140.174, hosts a RESTful API over the HTTP protocol, leveraging TCP for data transmission. It listens on ports 8081 and 8082 for different operations.
For encryption, OpenMV devices or smartphones upload images—either captured in real time or selected from the gallery—to the server via POST requests on port 8081. To retrieve and view encrypted images, the client sends a GET request to port 8082, including the encryption key in the request. The server then decrypts the corresponding image and returns the processed data.The test results are shown in \Cref{Fig.21}(a).

Case 2: Real-Time WiFi Transmission Encryption. This method employs a computer, a router, and three STM32 devices equipped with ESP8266 modules for secure real-time image transmission, as illustrated in \Cref{Fig.20}(b). The three STM32 devices function as a publisher, a correct key holder (authorized police personnel), and an incorrect key holder (a potential hacker). The host device reads confidential images from an SD card using the FAT file system, encrypts them, and stores them securely. It then connects to a WiFi network via the ESP8266 module, obtains the IP address 172.18.110.152, and establishes a listening service on port 8083 using the TCP protocol. When a client device requests access to image data, it sends a request to the host at IP:8083. If the requesting device provides the correct decryption key, it successfully retrieves and decrypts the image. However, an incorrect key results in receiving only the encrypted image, ensuring the confidentiality of law enforcement data and preventing unauthorized access.The test results are shown in \Cref{Fig.21}(b).
\section{Conclusion}
\label{sec5}
This paper introduces the MHDNN, constructed by coupling two heterogeneous one-dimensional neurons with a memristor acting as the synapse between them. The number and types of fixed points in the model are determined by the parameters and initial conditions, leading to complex dynamical behavior. Through various numerical simulation methods, such as bifurcation diagrams, attractor phase diagrams, and basin of attraction analysis, we investigated the influence of parameters and initial conditions on the system's dynamics. The results reveal a wide range of quasi-periodic, periodic, chaotic, and hyperchaotic behaviors, as well as diverse firing patterns and synchronization phenomena between the two neurons.  
Additionally, we designed a PRNG based on the MHDNN map, demonstrating high randomness in its output. An image encryption algorithm was also developed using the MHDNN model, with multiple evaluations confirming its security and robustness. Furthermore, we designed a multi-scenario PIoT image encryption platform, proving its practical value. Looking ahead, to advance encryption in multimodal integrated systems, we will further explore the algorithm’s applications in audio and text encryption and work toward its hardware implementation.
\bibliographystyle{elsarticle-num}
\bibliography{ref}
\end{document}